\documentclass[a4paper,fleqn,usenatbib,useAMS]{mnras}

\usepackage{graphicx}	
\usepackage{amsmath}	
\usepackage{amssymb}	
\usepackage{multicol}        
\usepackage{bm}		
\usepackage{pdflscape}	
\usepackage{placeins}
\usepackage{caption}
\usepackage{subcaption}

\usepackage[T1]{fontenc}
\usepackage{ae,aecompl}

\usepackage{newtxtext,newtxmath}

\title[Partial Disruption of White Dwarfs by Intermediate-Mass Black Holes in Eccentric Orbits]{Partial tidal disruptions of spinning eccentric white dwarfs by spinning intermediate-mass black holes}
	
\author[D. Garain and T. Sarkar]{
	Debojyoti Garain$^{1,2}$\thanks{E-mail: dgarain@clemson.edu}, 
	Tapobrata Sarkar$^{1}$\thanks{E-mail: tapo@iitk.ac.in}
	\\
	$^{1}$Department of Physics, Indian Institute of Technology Kanpur, Kanpur 208016, India\\
	$^{2}$Department of Physics and Astronomy, Clemson University, Clemson, SC 29634, USA
}

\date{Accepted XXX. Received YYY; in original form ZZZ}

\pubyear{2024}

\begin{document}
\label{firstpage}
\pagerange{\pageref{firstpage}--\pageref{lastpage}}
\maketitle

\begin{abstract}
Intermediate-mass black holes (IMBHs, $\sim 10^2-10^5M_{\odot}$) are often dubbed as the missing link between stellar mass ($\lesssim 10^2M_{\odot}$)
and super-massive ($\gtrsim 10^{5-6} M_{\odot}$) BHs. Observational signatures of these can result from tidal disruptions of white dwarfs (WDs), which 
would otherwise be captured as a whole by super-massive BHs. Recent observations indicate 
that IMBHs might be rapidly spinning, while it is also known that isolated white dwarfs might have large spins, with spin periods of the order of 
minutes. Here, we aim to understand the effects of ``coupling'' between BH and stellar spin, focussing on the tidal disruption of 
spinning WDs in the background of spinning IMBHs. 
Using smoothed particle hydrodynamics, we perform a suite of numerical simulations of partial tidal disruptions, where spinning WDs are in eccentric orbits about 
spinning IMBHs.  We take a hybrid approach, where we integrate the Kerr geodesic equations while being in a regime where we can treat 
the internal stellar fluid dynamics in the Newtonian limit. The coupling of BH and stellar spin results in distinctive 
behaviour of mass distribution of debris, compared to non-rotating cases. Further, while late time fallback rates of debris into the BH is unaffected by only
BH spin, these have noticeable deviations in the presence of stellar spin, in particular, this causes a broadening of the fallback curves at late times.
On the other hand, gravitational wave signatures are unaffected by stellar spin in the parameter regime that we consider.
\end{abstract}





\section{Introduction}
\label{sec1}

A star is tidally disrupted when its self gravity is overcome by the tidal field of a nearby compact object such
as a {black hole} (BH), see \cite{FrankRees, Rees}.
Based on the outcome of the interaction, such tidal disruption events (TDEs) can be broadly categorized into two classes, quantified by the strength of the encounter, i.e, the impact parameter, $\beta = r_t/r_p$, 
where $r_p$ is the pericenter distance of the orbit, and $r_t$ is the tidal radius. In deep encounters, where the pericenter distance lies within the tidal radius (i.e., $\beta \gtrsim 1$), 
the star undergoes full disruption, leading to the formation of tidal tails. The bound portion of the disrupted material is subsequently accreted by the black hole \citep{Rees, Evans1989}. 
On the other hand, in grazing encounters where the pericenter distance is outside the tidal radius (i.e., $\beta\lesssim1$), only the outer layers are stripped off from the star, giving rise to a 
self-gravitating core -- defined here as the inner bound material that remains intact and gravitationally bound after the encounter -- alongside extended tidal tails \citep{Manukian2013, Gafton2015, Banerjee2023}.
TDEs are important and interesting astrophysical phenomena, since the in-falling debris dissipates energy, and this causes a 
luminous flare \citep{Evans1989, Hayasaki2013, Hayasaki2016, Bonnerot2016, Liptai2019, Clerici2020}. Several such events have
been observed and are recorded in the literature see, e.g. \cite{Holoien}. Indeed, TDEs have been studied over many decades now, and several analytical
and numerical tools have been developed to date.
A remarkable fact about TDEs is that the fallback rate of debris for full disruptions at late times follows the ubiquitous $t^{-5/3}$ power law with time $t$,
as initially analytically derived by \cite{Rees,Phinney}, 
and further validated through numerical simulations by \cite{Lodato} for various stellar structure profiles, and subsequently confirmed in observed light curves \citep{Komossa, van}.
However, a number of unresolved issues exist, that point to the tension between theoretical models and experimental observations.
For example, as discussed in the review by \cite{LodatoRev}, the $t^{-5/3}$ behaviour has been seen in ultra-violet and optical lightcurves
of TDEs \citep{Gezari, Shu}, while \cite{LodatoRossi} argue that these should be much flatter and follow the power law $t^{-5/12}$ a few
months after the TDE. 
Also, as discussed in the recent review of \cite{GezariRev}, the observed radii of circularised debris disk after a TDE is often larger 
than the theoretically predicted values. Further, the theory of partial TDEs show substantial deviations from the power laws stated above, 
which have not been observationally confirmed as of now.
Extensive discussions on theoretical and observational aspects of TDEs appear in the book chapters of \cite{Jonker}. Further recent reviews that discuss various theoretical aspects of TDEs and related relativistic effects 
can be found in \cite{Stone2019, Weavers}. 

Astrophysical BHs are often classified according to their masses. 
Whereas several stellar mass BHs ($\sim 3 - 100 M_{\odot}$) \citep{Strader2012, Corral-Santana2016} and supermassive BHs ($\gtrsim 10^6 M_{\odot}$)
\citep{LauraHolland, Andika}
have been observed, observational evidence for intermediate-mass BHs (IMBHs) remains relatively rarer, and some of these have been recently 
reported in \cite{IMBH0, IMBH1, IMBH2, IMBH3, IMBH4}. Indeed, recent observations by \cite{Cao2023} has indicated that the BH 
that caused the TDE 3XMM J150052.0+015452 is a $\sim 10^5 M_{\odot}$ BH with the dimensionless spin parameter $a > 0.97$. 
Now, it is currently believed that IMBHs constitute the missing link in the theory of BH formation \citep{Volonteri}, see \cite{Green} for a recent review.
It is thus imperative to study IMBHs, in particular spinning ones (Kerr IMBHs) in further details, as theoretical results should 
be able to provide additional diagnostic indicators for these important class of objects. TDEs are important in this context, as these 
often provide ``smoking gun'' signatures of BHs. In particular, white dwarfs (WDs) with their well known equation of state are 
attractive candidates for such studies, as these can only be tidally disrupted by IMBHs \cite{Jonker}.

Numerical studies of tidal disruptions of stars by supermassive BHs continue to receive considerable attention, 
for a comprehensive review of the current literature, see the book \cite{Jonker}. 
Aspects of TDEs from stellar mass BHs have also been well studied, see e.g., the recent work of \cite{Wang2021, Vynatheya}. 
In contrast, studies of TDEs by IMBHs have been relatively fewer.
The work of \cite{Rosswog2009} is one of the first that studied numerical smoothed particle hydrodynamics (SPH) simulations of 
full TDEs involving static (Schwarzschild) IMBHs and non-spinning WDs, with the latter in a parabolic trajectory. 
More recently, \cite{Chen2023} and \cite{Claire} have studied partial disruptions in this system with the WD in an eccentric trajectory. 

Apart from the intrinsic importance of studying TDEs from IMBHs, the introduction of stellar rotation (spin)
in the background of spinning IMBHs is a further motivation of our study. Specifically, we aim to understand how stellar spin ``couples'' to the 
spin of Kerr IMBHs during a TDE. We will, in particular, study rapidly spinning WDs. Theoretical
and observational aspects of spinning WDs have been studied for many decades now, see e.g. \cite{ShapiroTeukolsky}, a more recent
review appears in \cite{Kawaler}.
Indeed, although many known WDs spin slowly with spin rotational periods of a few to several tens of hours \citep{Hermes}, rapidly spinning 
WDs (which we take to be ones with spin periods of the order of minutes) are also known in the literature.  \cite{Pelisoli2022} 
reported an accreting WD in the cataclysmic variable system J024048.51+195226.9 spinning with a spin period of $24.9$ 
seconds, while \cite{deOliveira} reported an accreting WD in such a system J2056-3014 with a $29.6$ second spin period. 
Rapidly spinning isolated WDs have also been reported in the literature. \cite{Kilic2021} reported on J2211+1136 which is an isolated
$1.2M_{\odot}$ WD with a spin period of $70$ seconds. In fact, \cite{Kawaler} estimates from scaling arguments that if one assumes conservation
of local angular momentum, a main sequence star of mass $\sim 3M_{\odot}$ that reaches a WD state with mass $\sim 0.6M_{\odot}$
may have a spin period as small as $\sim 100$ seconds. 
We also note that from a General Relativity (GR) perspective, \cite{Boshkayev} show that WD spin periods can indeed have theoretical lower bounds 
as small as fractions of a second. 

Given the fact that Kerr IMBHs can rotate close to extremality, and also emerging evidence for fast spinning
WDs, an astrophysical encounter between the two is certainly probable and interesting to study. 
Here, we complement and extend the existing analysis in the literature by considering a spinning WD and a Kerr IMBH, a viable astrophysical 
scenario which, to the best of our knowledge, has not been studied before in an SPH framework. 
In this paper we study a spinning $0.3M_{\odot}$ WD with a rotation period of $\sim 5$ minutes, which is
within the lower limit stated by \cite{Kawaler}, and in the background of a Kerr IMBH with a spin parameter $|a|=0.98$. Note that
typically, rapidly spinning WDs may harbour large magnetic fields. We will however exclude this from our analysis in this work. 

In the context of the above discussion, in a spinning WD - Kerr IMBH system, TDEs may provide crucial information regarding
their interactions, as the luminous flare from these depend on stellar properties, as well as on the mass and spin
of the central BH. These should be observationally interesting. 
The effects of different black hole masses on TDEs have been extensively investigated numerically in works such as \cite{RyuREl}, \cite{Wang2021}. Conversely, \cite{Gafton2019}, \cite{Jankovic} studied the effects of incorporating black hole spin on TDEs.
SPH based studies of TDEs involving spinning stars, on the other hand, is
a fairly recent addition to the literature, see \cite{Golightly2019b, Kagaya2019, Sacchi2019}. As pointed out by Rossi et. al. in page 16 
of article no. 40 of \cite{Jonker}, the reason behind this is perhaps the fact that tidal torques greatly spin up a star and initial 
stellar spin was thus thought to be of secondary importance unless the star initially rotates at speeds close to break-up.  
However, the works of \cite{Golightly2019b, Kagaya2019, Sacchi2019} indicate that this is probably not the full story, and that
there might be interesting effects due to stellar spin, such as steeper late time fallback rates \citep{Golightly2019b}, or failed 
disruption events \citep{Sacchi2019}.

There might indeed be several reasons for this. For example, the presence of stellar spin alters the structure of the star, causing it to deviate from spherical symmetry. 
Centrifugal force, along with self-gravity and pressure gradient, can become significant for spinning stars, resulting in an oblate spheroidal shape, 
with a larger equatorial diameter compared to the polar one. 
This may change the tidal radius. The theoretical estimate of the tidal radius for nonspinning stars is provided by the ubiquitous formula \citep{Hills, FrankRees,Lacy1,CarterLumineta,CarterLuminetb}, 
$r_t\sim R_{\star}(M/M_{\star})^{1/3}$, where $M$ is the mass of the BH, and $M_{\star}$ and $R_{\star}$ are the mass and radius of the star, respectively. 
For spinning stars, the tidal radius can differ compared to non-spinning stars due to the combined effects of centrifugal force, self-gravity, and tidal force,
and is a function of the direction of stellar spin as well. Studies such as \cite{Kesden2012spin} and \cite{Golightly2019b} have provided approximate formulas for tidal radius, showing that both the magnitude and direction of stellar spin can influence the tidal radius and, consequently, the strength of the interaction, $\beta = r_t/r_p$. For example, with fixed $r_p$, an increase in $\beta$ due to stellar spin will make the TDE into an effectively deeper encounter compared to one without such spin.

On the other hand, even if this change in $\beta$ can be ignored (see below), stellar spin alone might affect the result of a TDE, as tidal interactions induce a spin 
in the disrupted star via a tidal torque. With an initial prograde stellar spin (where the stellar rotation axis is parallel to the orbital angular momentum
of the star), the tidal torque, whose effect is maximal near the pericenter, causes a spin-up of the star, due to which outer layers
of the star are more susceptible towards disruption than in the non-spinning case as a result of increased centrifugal force. On the other hand, with 
retrograde spin (where the stellar rotation axis is antiparallel to the orbital angular momentum
of the star), the tidal torque first spins
the star down, and then causes a spin-up in the prograde sense, so that overall the star is more resistant to disruption compared to the prograde spin case. See the 
discussion in \cite{Golightly2019b, Sacchi2019}. 

In this paper, we use SPH to perform a numerical study of partial TDEs involving spinning white dwarfs (WDs)
that are in eccentric orbits in an intermediate-mass (spinning) Kerr BH background. The numerical algorithm used in this study has been developed
by us \citep{Banerjee2023, Garain2023a, Garain2023b} and extensively tested against many available results that often use the popular publicly available
code PHANTOM \citep{Price2018} which inspired an initial version of our code. 
To facilitate comparison and demonstrate the reliability of our implementation, we include a representative set of standard test results in the Appendix~\ref{app:codetest}. These tests validate the key components of our SPH code -- including hydrodynamics, self-gravity, external forces, and the tidal disruption setup -- against known analytical solutions and established benchmarks.

Here, we have chosen stellar parameters so that the change in the impact parameter of the tidal interaction is negligible, so that we can comment on effects
attributed solely to initial stellar spin with reasonable confidence. In particular, along with the black hole mass $M = 10^4M_{\odot}$, 
we choose the mass of the WD to be $M_{\rm wd} = 0.3M_{\odot}$ and hence its radius $R_{\rm wd}=0.018R_{\odot}$.
This follows from the WD equation of state (EOS), considering the WD to be a degenerate electron gas at zero temperature  
\citep{ShapiroTeukolsky}, given in parametric form by
\begin{eqnarray}
P  &=& K_{P} \Bigl[x(1+x^2)^{1/2}(2x^2/3-1)+ \log_{e}\bigl[x+(1+x^2)^{1/2}\bigr] \Bigr]~,\nonumber\\
\rho  &=&  K_{\rho} x^3~,
\end{eqnarray}
where the constants $K_{P} = 1.4218\times10^{24}/(8\pi^2)$  $\rm N~m^{-2}$, $K_{\rho} = 1.9479 \times 10^9$ $\rm kg~m^{-3}$, 
and $x=p_F/(m_e c)$ is the non-dimensionalised Fermi momentum with $p_F$ the Fermi momentum of the 
electron gas, $m_e$ the electron mass, and $c$ is the speed of light.

We checked that for this WD mass with a spin period $\sim 5$ minutes, 
the maximal ratio of the equatorial radius to the polar radius before disruption is $\sim 1.015$. 
Additionally, for this spin period, the volume equivalent radius for the spinning WD differs by approximately $0.5\%$ from its
non-spinning counterpart, confirming that the spherical shape is largely preserved. Furthermore, using the tidal radius formula provided by \cite{Golightly2019b}, the tidal radius for this spinning WD changes by less than $1\%$ compared to a non-spinning WD. This indicates the fact that here, the change in $\beta$ has a negligible effect while fixing $r_p$, so that we can study effects solely due to stellar spin.
\footnote{We also note in passing that akin to rotation, possible modifications to GR that affects stellar radius for a given stellar mass also has the same effect, as discussed in \cite{Garain2023a}.}

In the backdrop of the above discussions, the rest of the paper focuses on TDEs of spinning WDs in the background of spinning IMBHs. 
The next section \ref{sec2} explains the methodology followed in this paper; section \ref{sec3} details our main results.
Specifically, we  study the fate of bound debris and fallback rates after a partial tidal encounter of the spinning WD described
above with a Kerr IMBH. Next, we discuss the spin properties of the surviving core and also compute gravitational wave amplitudes 
for the TDEs. Finally, section \ref{sec4} ends this paper with discussions and conclusions, after we discuss a few conceptual
issues associated with the assumptions made in this paper.  Two appendices \ref{app:convergence} and \ref{app:m1m2}
are provided to substantiate some the numerical results presented in the paper. 

\section{Methodology}
\label{sec2}

\subsection{Incorporating Spinning Black Hole and Spinning Star in SPH Code}
\label{sec2a}

In this section, we outline our methodology for numerically simulating the tidal disruption of a spinning WD interacting with a spinning BH. Our approach involves modifying the SPH code developed by \cite{Banerjee2023} to account for the gravitational influence of the spinning BH.
To incorporate relativistic effects, we adopt a hybrid approach, integrating the exact relativistic acceleration due to the spinning BH with a 
Newtonian treatment of hydrodynamics and self-gravity. This is certainly an approximation, but as we will discuss in the final section \ref{sec4}, gives 
fairly accurate results given our choice of BH and stellar parameters. 

The Kerr metric, expressing the geometry of spacetime around a spinning BH in Boyer-Lindquist coordinates ($t, r, \theta, \phi$), is employed to calculate the acceleration due to Kerr spacetime. The metric is given by :
\begin{eqnarray}
	\nonumber
	ds^2 = -\left(1-\frac{2 G M r}{c^2\,\Sigma}\right)dt^2 - \frac{4 G M r a \sin^2\theta}{c\,\Sigma}dtd\phi + \frac{\Sigma}{\Delta}dr^2\\ + \Sigma d\theta^2
	+ \left(r^2 + a^2 + \frac{2 G M r a^2 \sin^2\theta}{c^2\,\Sigma}\right)\sin^2\theta d\phi^2
	\label{Eq.metric}
\end{eqnarray}
where $\Sigma = r^2 + a^2\cos^2\theta$, $\Delta = r^2 +a^2 - 2 G M r/c^2$, and $a = J/M c$ represents the Kerr parameter, with $J$ and $M$ being the angular momentum and mass of the BH, respectively, and $G$ is Newton's gravitational constant. The dimensionless spin parameter $a^{\star} = (c^2\,a)/(M\,G)$ ranges 
from $-1$ to $+1$, where $a^{\star}>0$ indicates prograde orbits and $a^{\star}<0$ indicates retrograde orbits.
We note that although many authors in the literature use Boyer-Lindquist coordinates as we have done here, 
one could also have employed Kerr-Schild coordinates for describing the Kerr BH. A pleasant feature of these latter coordinates
is that there is no coordinate singularity at the event horizons. Our usage of Boyer-Lindquist coordinates is justified by the fact that we
are well away from the event horizons of the Kerr BH, and these coordinates should be robust in the regions of our interest since
we are not studying accretion phenomena. We
point out that \cite{Tejeda2017} made a detailed comparison between simulation results involving Boyer-Lindquist and 
Kerr-Schild coordinates in some situations involving close stellar encounters with SMBHs and found little difference between the two.

To combine the external acceleration from the Kerr BH with the SPH acceleration, we use the geodesic equation expressed in global coordinate time $t$:
\begin{equation}
	\frac{d^2 x^i}{d t^2}  = -\left(g^{i\sigma}-\dot{x}^i g^{0\sigma}\right)\Bigg( \frac{\partial g_{\gamma\sigma}}{\partial x^{\delta}} -\frac{1}{2}\frac{\partial g_{\gamma\delta}}{\partial x^{\sigma}} \Bigg)\dot{x}^{\gamma}\dot{x}^{\delta},
	\label{Eq.geodesic}
\end{equation}
where $i=1,2,3$ (the spatial coordinates), Greek indices range from 0 to 3, and an overdot indicates a derivative with respect to coordinate time.
Equation \ref{Eq.geodesic} can be derived concisely from the standard geodesic equation expressed in terms of proper time $\tau$ \citep{Misner1973}:
\begin{equation}
	\frac{d^2 x^{\mu}}{d \tau^2} + \Gamma_{\gamma\delta}^{\mu}\frac{d x^{\gamma}}{d \tau}\frac{d x^{\delta}}{d \tau} = 0,
\end{equation}
with Christoffel symbols:
\begin{equation}
	\Gamma_{\gamma\delta}^{\mu} = \frac{1}{2}g^{\mu\sigma}\left(\partial_{\delta} g_{\gamma\sigma} + \partial_{\gamma} g_{\delta\sigma} - \partial_{\sigma} g_{\gamma\delta}\right).
\end{equation}
Expressing the derivatives with respect to coordinate time $t$, one obtains:
\begin{equation}
	\frac{d^2 x^\mu}{d\tau^2}= \frac{d}{dt}\left[\frac{dx^\mu}{dt}\Lambda\right]\Lambda,
\end{equation}
where $\Lambda=dt/d\tau$. Substituting this into the geodesic equation and rearranging terms yields:
\begin{equation}
	\ddot{x}^{\mu}+ \dot{x}^{\mu} \frac{d^2t}{d\tau^2}\left(\frac{d\tau}{dt}\right)^2 = -\Gamma_{\gamma\delta}^{\mu}\dot{x}^{\gamma}\dot{x}^{\delta}.
\end{equation}
For $\mu=0$, noting $\ddot{x}^{0}=0$ and $\dot{x}^{0}=1$, we get:
\begin{equation}
	\frac{d^2t}{d\tau^2}\left(\frac{d\tau}{dt}\right)^2=-\Gamma^0_{\gamma\delta}\dot{x}^{\gamma}\dot{x}^{\delta}.
\label{Eq.geodesic1}
\end{equation}
Inserting this back into the spatial components ($\mu=i$) immediately provides Equation \ref{Eq.geodesic}.

There is an important issue that needs to be addressed here. It is well known that spinning particles do not exactly follow geodesics in Kerr BH backgrounds and one
should more appropriately use the Mathisson-Papapetrou formalism \citep{Matthison, Papapetrou, Semerak}. We can however estimate the deviation from
geodesic motion following \cite{Semerak}, and find that it is small in the situation that we consider (see section \ref{sec4}). Hence, to a very good approximation, we can use
the Kerr geodesic equations to estimate the effects of the Kerr IMBH.    

As our numerical code appropriately uses Cartesian coordinates, we express the metric in Equation \ref{Eq.metric} in Cartesian-like coordinates 
($t,x,y,z$), related to ($t,r,\theta,\phi$) as:
\begin{equation}
	x = \sqrt{r^2 + a^2}\sin\theta\cos\phi~,~~
	y = \sqrt{r^2 + a^2}\sin\theta\sin\phi~,~~
	z = r\cos\theta.
\end{equation}
Next, using the contravariant metric components and analytically differentiating the covariant metric components in Cartesian-like coordinates, 
we calculate the acceleration in $x, y, z$ for all SPH particles using Equation \ref{Eq.geodesic}.

To verify our implementation, we perform two key tests. First, we integrate Equation \ref{Eq.geodesic} for a point mass particle using a 
fourth-order Runge–Kutta integrator (RK4) independent of our SPH code, and compare the results with the geodesics from \cite{LiptaiGRSPH} (their Figure 16). 
This initial verification ensures the correctness of our test particle geodesic. 

\begin{figure*}
	\centering 
	\includegraphics[scale=0.45]{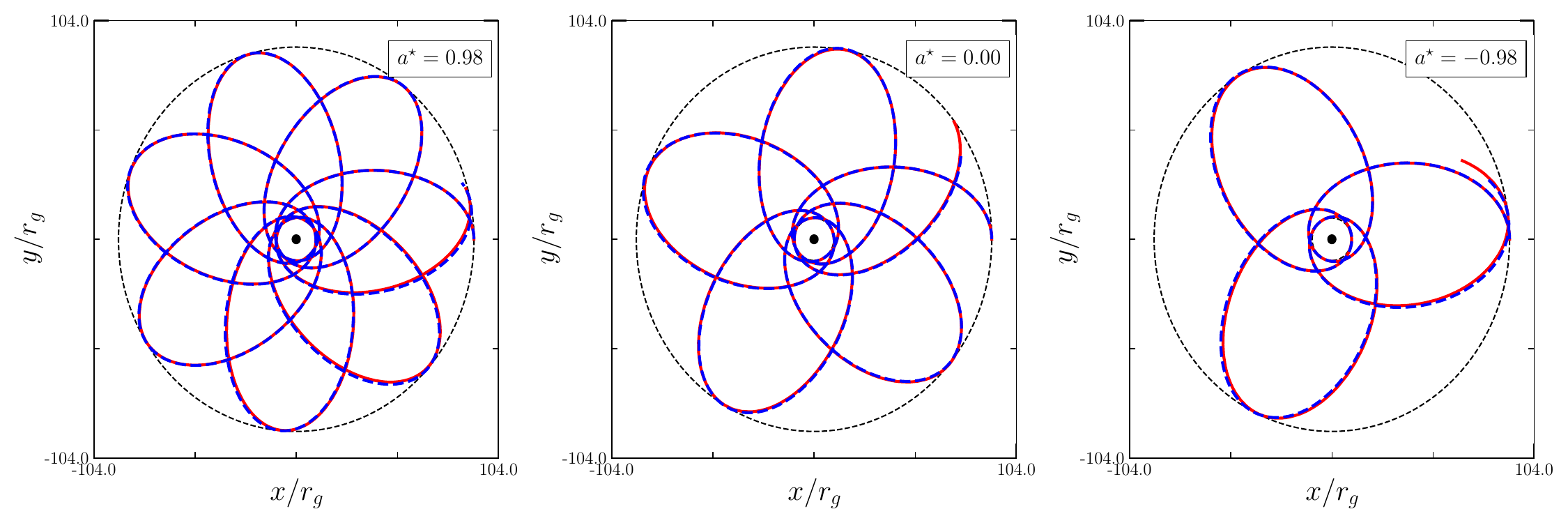}
	\caption{This figure shows the trajectories of a rigid body in Kerr space-time for various spin parameter values, as indicated in the legend. The center of mass of the rigid body, obtained from SPH simulations, is represented in red. Additionally, a dashed blue line depicts the trajectory for the test particle geodesic, obtained independently of SPH. In all SPH simulations, the center of mass is situated on the equatorial plane. The apocenter ($r_a$) and pericenter ($r_p$) remain constant across all simulations, with the outer black dotted circle denoting the apocenter.}
	\label{fig.BHspin}
\end{figure*}

To ensure correct implementation in the SPH code, we perform long-duration simulations of the orbital motion of a rigid body assuming a polytropic EOS around a $10^4 M_{\sun}$ Kerr BH in the equatorial plane. For this purpose, we create a spherical 1 $M_{\sun}$ rigid body with a radius chosen such that $r_p\approx 10r_g$, where $r_g = GM/c^2$ is 
the gravitational radius of the BH, where, following the discussion of \cite{Stone2019}, we assume that it is still reasonable to treat the stellar hydrodynamics in the Newtonian limit for this choice of $r_p$.
In these simulations, the pericenter distance is significantly greater than the tidal radius, ensuring that the star is not tidally interacting and maintaining its spherical shape upon passing the pericenter.  We then verify that the stellar center of mass follows the point particle geodesic obtained independently of the SPH code with the same initial conditions for different $a^{\star}$ values, see Figure \ref{fig.BHspin}. This test verifies the accuracy of the Kerr metric acceleration within our SPH code.

Next, to create a uniformly spinning star in a relaxed state, we follow the procedure outlined in \cite{Garc2020}. Initially, we generate a non-rotating fluid star in SPH using equal-mass particles arranged in a close-packed sphere. Their radial positions are then adjusted using the well-known stretch mapping technique introduced by \cite{Herant}, which ensures that the final particle distribution follows the desired density profile. 
Specifically, a particle initially located at radius $r_0$ is shifted to a new radius $r$ such that
\begin{equation}
	\frac{M_{_{LE}}(r)}{M_{_{LE}}(r_{max})}=\frac{r_0^3}{r_{max}^3},
\end{equation}
where $M_{{ LE}}(r)$ is the enclosed mass from the desired Lane–Emden profile, and $r_{\rm max}$ is the radius of the initial close-packed sphere.
For polytropic models, the density profile is derived from the Lane–Emden equation, while for WD models, it is obtained by solving the hydrostatic equilibrium and mass conservation equations, as detailed in \cite{Garain2023b}. This stretching process reduces noise by aligning the particle distribution more accurately with the target density, in contrast to methods that use random particle placement and suffer from Poisson noise. 
During the subsequent relaxation phase in isolation, residual motions are damped by the standard Monaghan–Gingold artificial viscosity \citep{Morris}, along with the Balsara switch \citep{Balsara1995} to suppress viscosity in shear flows.
The final relaxed fluid star is attained once the kinetic energy falls below a specified threshold and the density profile converges.

\begin{figure*}
	\centering 
	\includegraphics[scale=0.35]{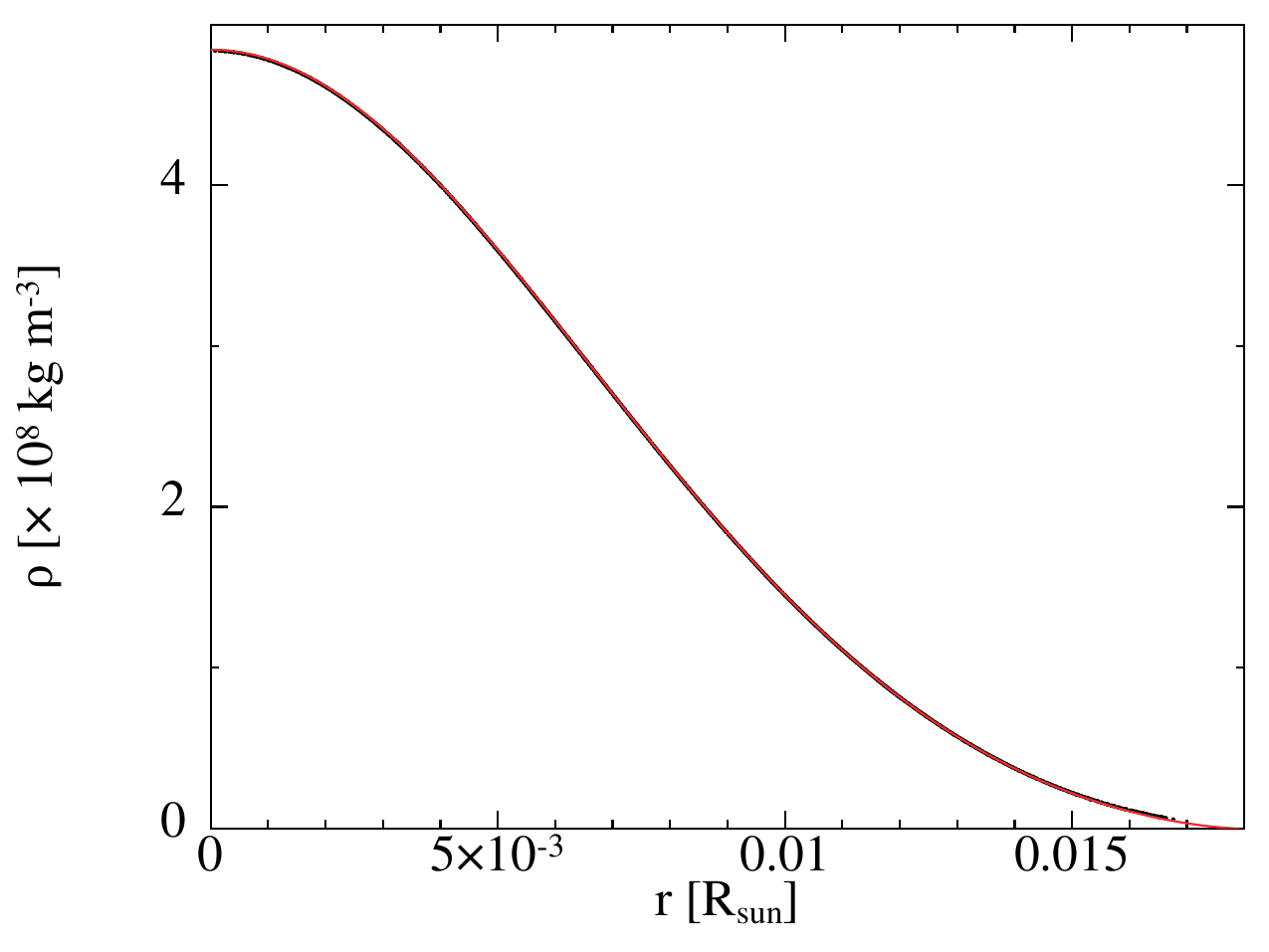}
	\includegraphics[scale=0.35]{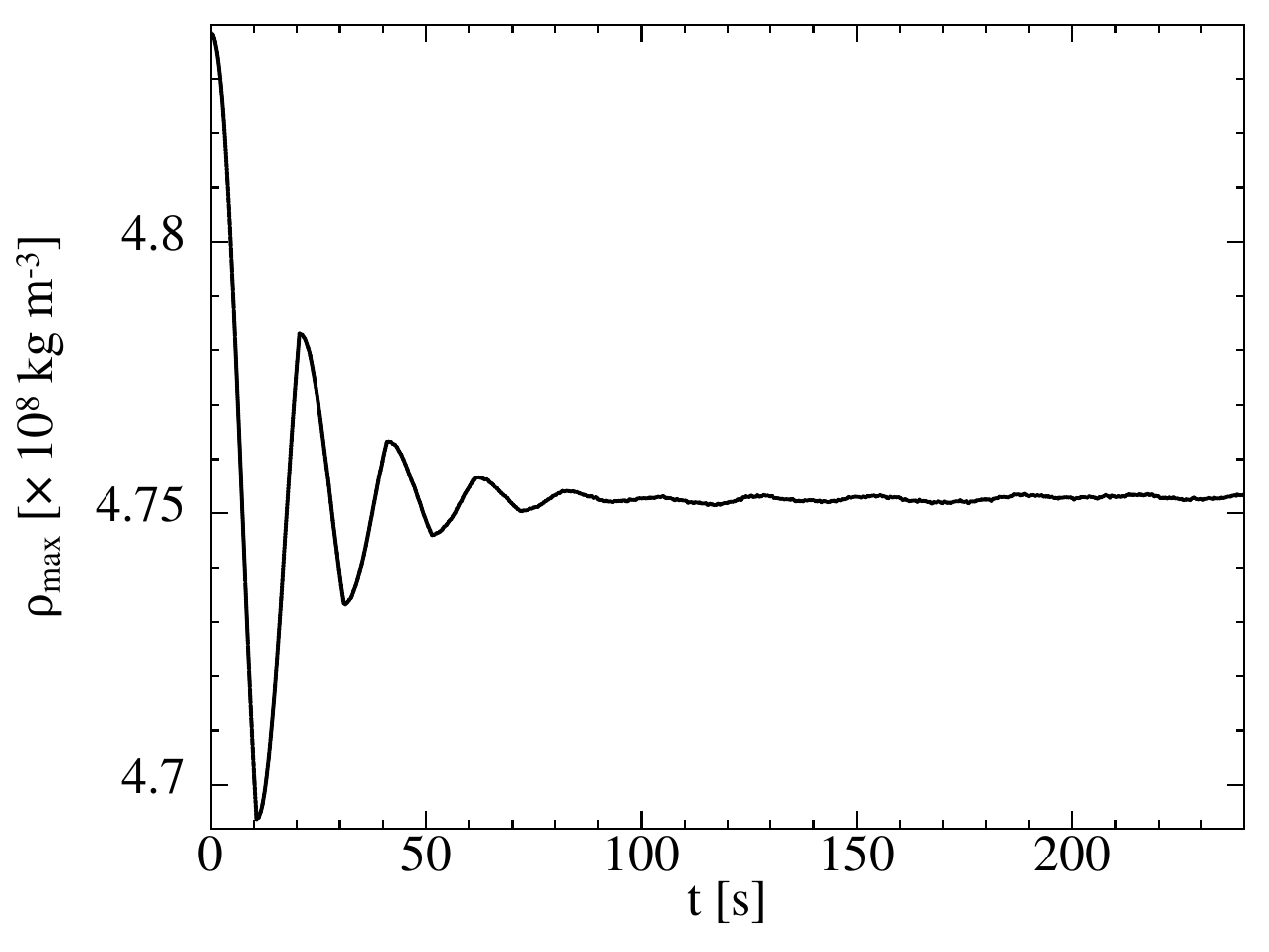}
	\includegraphics[scale=0.35]{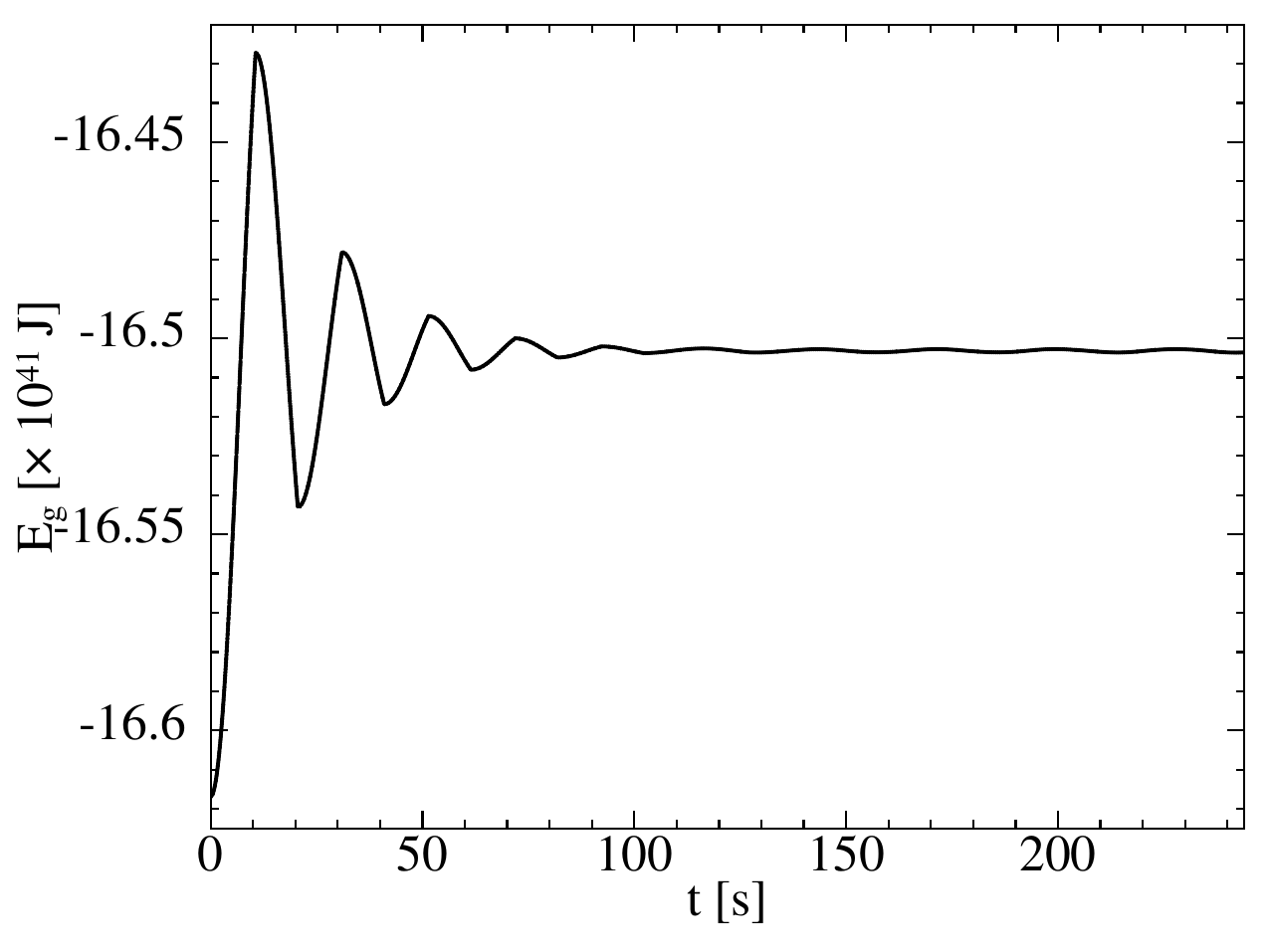}
	\includegraphics[scale=0.35]{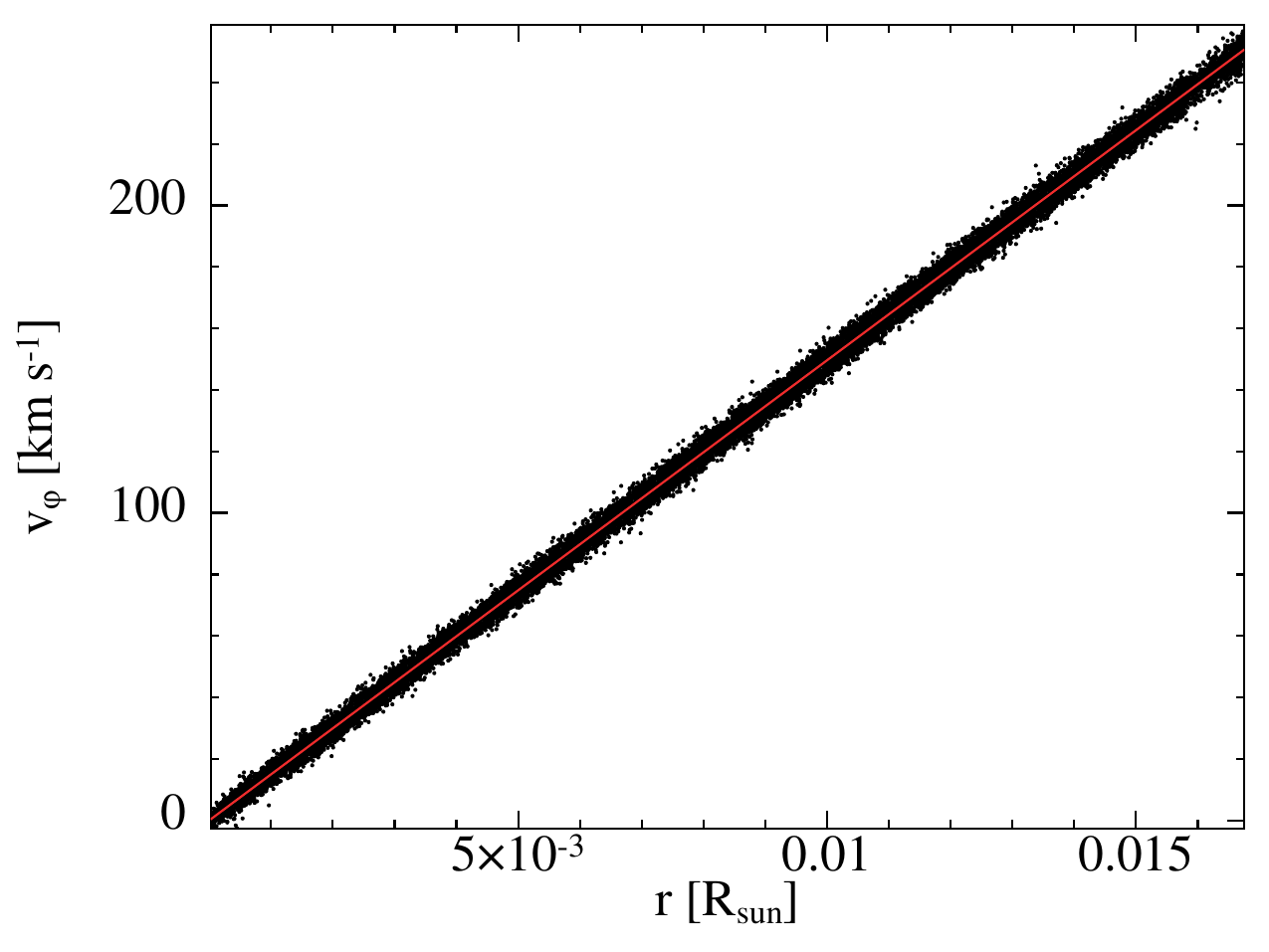}
	\caption{\textbf{Top left:}  The density is plotted against the radius for SPH particles using black dots for a relaxed, non-spinning WD of mass $0.3 M_{\odot}$, with the theoretical solution depicted by the red line. \textbf{Top right:} The maximum central density is depicted over time after imposing rigid rotation. \textbf{Bottom left:} The total self-gravitational energy is plotted with time. \textbf{Bottom right:} The azimuthal velocity for SPH particles (black dots) plotted against distance from the rotation axis, alongside the theoretical solution (red line) after relaxation. These figures are generated using SPLASH \citep{Price2007}.}
	\label{fig.rot}
\end{figure*}

Once we obtain the relaxed fluid star without rotation, we introduce rigid rotation and allow the system to relax in the presence of internal dissipation, i.e., artificial viscosity. To eliminate velocity field noise and quickly reach equilibrium, we periodically set velocities to zero in the co-moving reference frame. During the initial relaxation, resetting occurs at shorter intervals, $dt = t_{sc}/3$ (where $t_{sc}$ is the sound crossing time), repeated five times consecutively. After that, the system is reset at longer intervals, $dt = 0.8t_{sc}$, until the fluctuation in angular velocity is $0.1\%$ of the desired angular velocity for star rotation. Finally, we allow the system to evolve freely without resetting velocities for a sufficiently long period, ensuring oscillations around average values of star properties -- central density, polar and equatorial radius, and total kinetic, internal, and gravitational energies -- remain below $2.5\%$. This methodology has been validated as widely applicable to various stellar models, ranging from polytropic stars to WDs.

The maximum spin that a star can sustain without breaking apart is determined by setting the self-gravitational force of the star equal to the centrifugal force. The angular velocity at which the star would break up is then calculated as $\Omega_{\rm br} = \sqrt\frac{G M_{\star}}{R_{\star}^3}$. Here, $M_{\star}$ and $R_{\star}$ represent the mass and radius of the star, respectively. If the star is spinning with an angular velocity $\Omega_{\star}$, the break-up fraction is defined as $\lambda=\Omega_{\star}/\Omega_{\rm br}$. 
We note that the above analysis is done assuming that the star is spherical, so that $\Omega_{\rm br}$
is the break-up angular speed at the stellar equator. If the star on the other hand is assumed to be spheroidal, then by using the Roche model, the maximum value of 
the break up angular speed can be calculated to be $\sim 0.54\Omega_{\rm br}$ \citep{ShapiroTeukolsky}. In this paper, we will conservatively choose $\lambda=0.15$.

In Figure \ref{fig.rot}, we present the process of creating a relaxed, uniformly spinning WD with a mass of $0.3 M_{\odot}$ and a radius of $0.018 R_{\odot}$. In the top left panel, we show the density profile of the SPH particles after the initial relaxation from the stretch map, demonstrating a close match with the theoretical profile obtained independently of the SPH code. After achieving the relaxed profile for the non-spinning WD, we set up the WD in rigid rotation with an angular velocity of $\Omega_{\star} = \lambda\,\Omega_{\rm br}$, specifically choosing $\lambda=0.15$. Following a few velocity resets in the co-moving reference frame, we allow the spinning WD to evolve over a significant period, observing that fluctuations in its properties remain within acceptable tolerance levels. In particular, we show the variation of two properties, maximum central density $\rho_{\rm max}$ and total self-gravitational energy $E_{\rm g}$, as functions of time in the top right and bottom left panels of Figure \ref{fig.rot}, respectively. In the bottom right panel, we present the azimuthal velocities of the particles as a function of the distance from the rotational axis after the relaxation of the spinning WD, along with the exact solution ($v_{\phi} = \Omega_{\star} r$, where $r$ is the distance from the rotation axis). The close match between the numerical and theoretical solutions validates the correctness of our process for creating a uniformly spinning WD.

The rotation imposed on our WD corresponds to a rotational period of $\sim$ 5 minutes. This should be contrasted with the values obtained by 
\cite{Pelisoli2022} (24.93 seconds) and \cite{Kilic2021} (70 seconds) that we have mentioned in the introduction. Here, we consider the angular speed 
of the WD to be substantially less than the values reported in these works, which justifies to some extent our exclusion of magnetic field effects and we
focus on obtaining the necessary features for TDEs that are dependent on stellar spin. The issue of TDEs involving spinning magnetic WDs is left for the future. 

\subsection{Simulation parameters}

To simulate TDEs, we employ the SPH code developed by \cite{Banerjee2023} with the modifications mentioned above. The code uses a fast recursive binary tree to efficiently search neighboring particles and calculate particle accelerations. The tree accuracy parameter, which determines when a distinct node acts as a multipole source of gravity or is further bisected, is set to $\theta = 0.5$. Standard artificial viscosity parameters, $\alpha^{\rm {AV}} = 1.0$ and $\beta^{\rm {AV}} = 2.0$, are used, and additionally, the Balsara switch is employed to reduce viscosity in shear flows \citep{Balsara1995}. The system evolves using a global time step, and SPH equations are integrated using the leapfrog integrator (K-D-K approach) to ensure conservation of mass, momentum, and energy.

To gain a comprehensive understanding of TDEs, works in the available literature have extensively explored their dependence on parameters such as $M$, $M_{\star}$, $R_{\star}$, the EOS of the star, eccentricity $e$ of the orbit, and impact parameter $\beta = r_t/r_p$. Some studies, such as those conducted by \cite{Tejeda2017}, \cite{Gafton2019}, have examined the impact of BH spin on TDEs and debris properties. On the other hand, studies including \cite{Kagaya2019}, \cite{Golightly2019b}, \cite{Sacchi2019}, have explored TDEs in the presence of stellar rotation. Notably, \cite{Golightly2019b} investigated the impact of stellar rotation on the fallback rate in the presence of a supermassive BH. However, to our knowledge, there has been no study examining TDEs by coupling BH spin with stellar rotation. 

To focus on the essential effects arising from the coupling of BH-WD rotations while maintaining a manageable parameter space, we initially fix the orbit, specifying the eccentricity and $\beta$.
We also include a spinning IMBH with a fixed mass and then set the mass, radius, and EOS for the WD. This narrows down the parameters of the system, 
leaving us with two independent variables corresponding to BH rotation and WD rotation.  We will then study how these rotations impact the observables. 
We remind the reader that we opt for an IMBH with a mass of $M = 10^4 M_{\odot}$ and a WD of mass $M_{\rm wd} = 0.3 M_{\odot}$ and radius $R_{\rm wd} = 0.018 R_{\odot}$. To relate the pressure to the density of the WD's fluid particles, we utilize a zero-temperature EOS, as described in \cite{Garain2023b}. We generate relaxed WDs using the methodology discussed in the preceding subsection (Section \ref{sec2a}), varying the break-up fraction ($\lambda$) with values of $0$, $0.05$, and $0.15$, where $\lambda = 0$ corresponds to a non-spinning WD. For BH rotation, we choose specific spin parameters, $a^{\star} = 0.98, 0.0, -0.98$, with $a^{\star} = 0$ corresponding to a Schwarzschild BH. Opting for higher values of the BH spin parameter aims to maximize rotational effects on observables. Subsequently, we consider an elliptical orbit with an eccentricity of $0.9$ and $r_p \simeq 39\,r_g$, where, as before, $r_g = G M/c^2$ represents the gravitational radius of the BH. This choice of $r_p$ ensures partial disruption. The presence of the surviving core modifies the late-time slope from the full disruption scenario \cite{Cufari2022}. The center of mass for both the relaxed spinning and non-spinning WDs is positioned initially at a distance of $5r_t \simeq 136\,r_g$ from the BH in the equatorial plane, with the BH located at the origin.

We simulate a single interaction of the WDs with the IMBH, and after this interaction, the tidally disrupted material forms two tails along with the surviving core, returning to the pericenter due to the initial bound orbit. Simulations continue until all material bound to the BH results after the first interaction is accreted. To manage computational costs, circularization and disk formation processes are omitted. The accretion radius is expanded from $2\,r_g$ to $3\,r_t \simeq 82 \,r_g$ as the bound material returns to the IMBH. 
The mass falling back onto the accretion radius is tracked over time, and its numerical derivative provides the fallback rate. For this differentiation, we apply first-order centred finite differences at interior points and forward/backward differences at the endpoints.
This way of calculating fallback rates directly from the simulations is outlined in \cite{Coughlin2015}, \cite{Golightly2019a}, \cite{Milesetal}, \cite{Garain2023a}. At this point, it is important to mention the work of \cite{Chen2023}, where the authors simulate the single tidal interactions between WD and IMBH with incoming eccentric orbits. They calculated the fallback rates using the well-known ``frozen-in'' approximation \citep{Lacy1, Rees,Lodato}. However, in this work, we rely on computing the fallback rates from the simulations, as it has been shown by \cite{Cufari2022} that self-gravity may play an important role in determining the structure of the debris. 

Once the WD interacts with the BH partially, its outer layers are stripped off, and the remaining portion of the WD contracts due to its self-gravity, forming a bound core. 
We identify this core using the energy-based iterative method of \citet{Guillochon2013}, also adopted by \citet{Banerjee2023}. Specifically, we begin by selecting the particle with the highest density and adopt its velocity, $\textbf{v}_{\text{peak}}$, as the initial estimate of the core velocity. In this frame, the specific binding energy of each particle is computed as
\begin{equation}
	\varepsilon_i = \frac{1}{2} \left(\textbf{v}_i - \textbf{v}_{\text{peak}}\right)^2 - \phi_i,
\end{equation}
where $\phi_i$ is the self-gravitational potential. Particles with $\varepsilon_i < 0$ are considered bound. We then compute the center-of-mass velocity of the bound set,
\begin{equation}
	\textbf{v}_{\text{core}} = \frac{\sum_{\varepsilon_i<0} \textbf{v}_i m_i}{\sum_{\varepsilon_i<0} m_i},
\end{equation}
and iterate the procedure until convergence. The final set of bound particles defines the core, from which we compute the bound core mass $m_{\text{core}}$ and center-of-mass position $\textbf{r}_{\text{core}}$. This method avoids arbitrary density thresholds or Roche approximations, providing a physically motivated definition of the remnant core.

The core results in the formation of a high-density region, which reduces the time step for the system's evolution. This poses a computational challenge for the system's long-term evolution. Hence, once the surviving core has moved a substantial distance away from the black hole and its properties have saturated, we replace the core particles with a sink particle. The detailed procedure for introducing the sink particle follows the methodology outlined in \cite{Garain2023a,Garain2023b}. We verified this procedure by changing the sink 
particle placement time to several different values, and observed no noticeable change in the fallback rate.


We employ $5\times 10^5$ SPH particles in our fiducial WD models. Convergence was verified by re-running representative cases with $5\times 10^6$ particles -- a tenfold increase -- which produced essentially identical results, confirming that the baseline resolution is adequate for the present study. The figures demonstrating this convergence test are included in Appendix~\ref{app:convergence}.


\section{Results}
\label{sec3}

In this section, we study the impact of rotation on the key observables of TDEs. All simulations start from identical initial positions of the WDs, and the pericenter distance 
$r_p$ is kept fixed. After the interaction, a self-gravitating core forms along with two tidal tails.
Once the core is identified, we exclude its particles and classify the remaining debris into leading and trailing tails based on the sign of the radial velocity in the orbital plane relative to the core's center-of-mass; their respective masses are then obtained by summing the constituent particle masses.
We look at the mass of the self-gravitating core, the fate of the bound debris, the rates at which material falls back, the spin of the core, and the emission of gravitational waves during the interaction. 
As we discuss below, the behaviour of these quantities for partial disruptions in elliptical orbits is notably distinct from ones in parabolic orbits.

Before we explore how the rotations of both the BH and the WD together affect the entire process, we will
first examine the effects of each rotation separately. This will help us get a clearer understanding of how BH rotation, stellar rotation, and their coupled effect 
influences various physical parameters.

\subsection{Fate of the Bound Debris and Fallback Rates}

\begin{figure*}
	\centering 
	\includegraphics[scale=0.25]{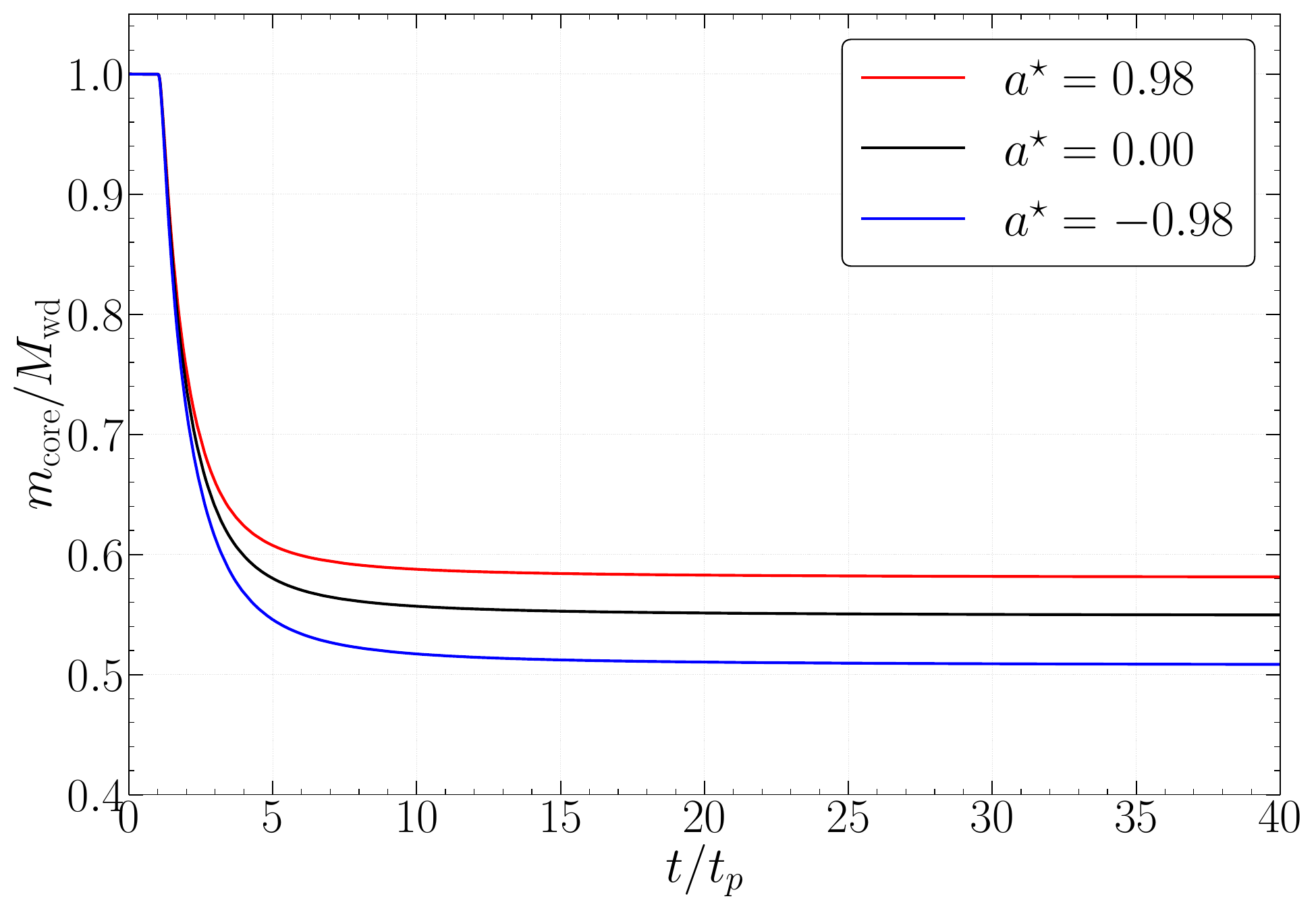}
	\includegraphics[scale=0.25]{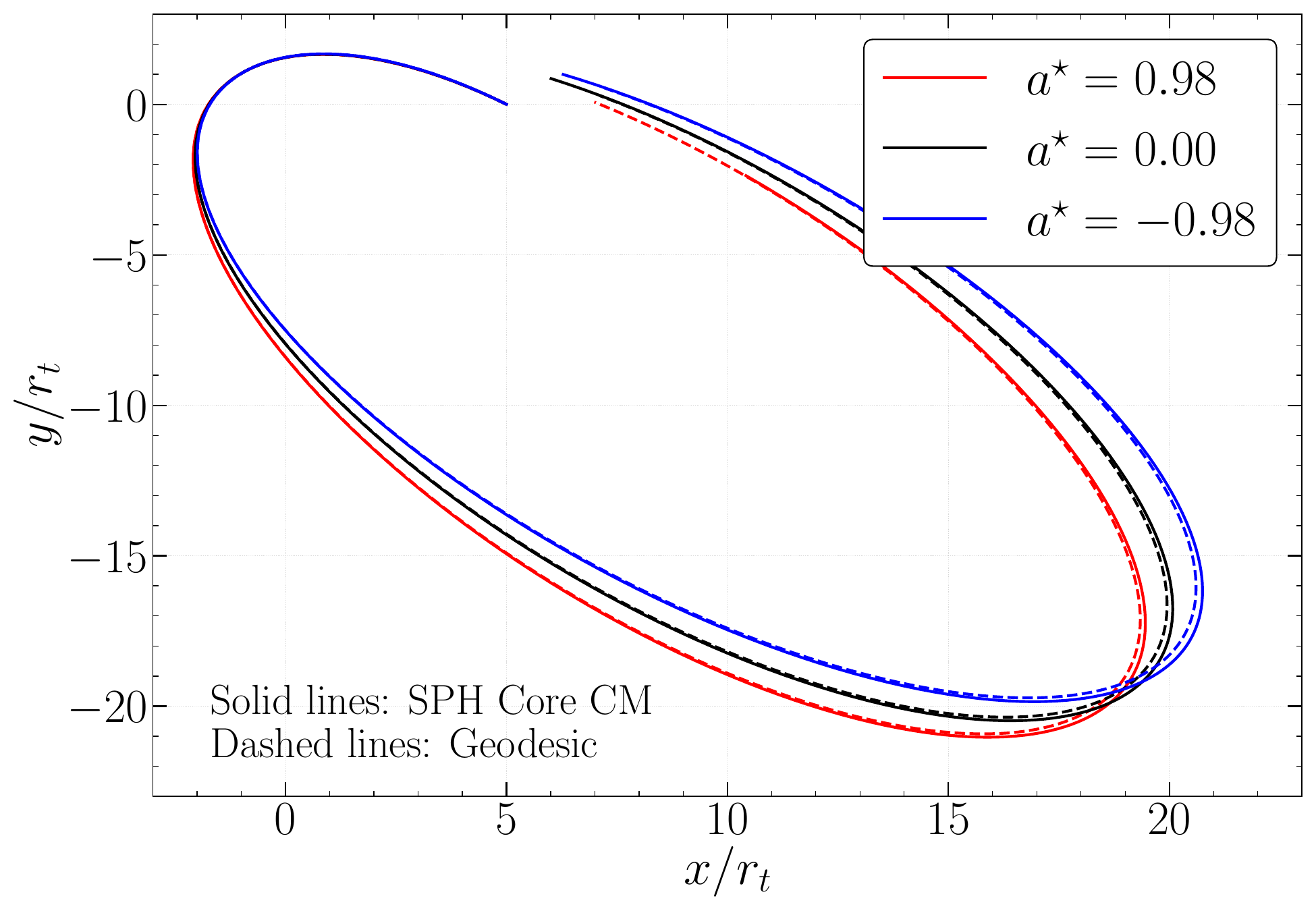}		
	\caption{\textbf{Left Panel:} Variation of the core mass fraction, $m_{\rm core}/M_{\rm wd}$, plotted against normalized time. \textbf{Right Panel:} Trajectory of the center-of-mass of the core plotted (solid lines) along with the test particle geodesics (dashed lines). The $x$ and $y$ axes are normalized with respect to the tidal radius ($r_t$).}
	\label{fig.RBH}
\end{figure*}

To determine the influence of BH rotation on the core and bound debris, we examine the tidal interaction between a non-spinning WD and a Kerr BH in elliptical trajectories. The core mass fraction, defined as the ratio of core mass over the initial mass, remains at $1.00$ until it reaches the pericenter. As a result of the WD's partial disruption, the core mass fraction decreases over time, finally settling to a constant value. However, as depicted in Figure \ref{fig.RBH}, left panel, the core mass fraction exhibits variation with the BH spin parameter $a^\star$. Retrograde orbits ($a^\star < 0$) experience more disruption compared to prograde orbits ($a^\star > 0$). Here, the time is normalized with respect to the time to reach the pericenter, denoted as $t_p$. Similar effects have been observed by \cite{Kesden2012} and \cite{Haas2012}. For the same impact parameter, they found that stars are more easily disrupted in retrograde orbits.

One possible reason for the increased disruption in a retrograde orbit is the longer time the WD spends near the pericenter, leading to more significant tidal interactions and greater mass loss compared to its prograde counterpart. This extended interaction, influenced by the larger apsidal precession angle in the retrograde orbit, enhances the likelihood of the WD experiencing stronger tidal forces.
We observe that as the WD approaches and interacts strongly with the BH near the pericenter, disruption begins and continues until it reaches the apocenter; eventually, the tidal tail masses converge. During this phase, we see from the right panel of Figure \ref{fig.RBH} that the retrograde orbit is closer to the black hole than the prograde orbit. This proximity leads to more pronounced tidal interactions, hence allowing tidal forces to overcome the WD's self-gravity to a greater extent. As a result, there is more disruption compared to the prograde orbit.
The precession of the orbits of the core center of mass for different $a^\star$ values is shown in the right panel of Figure \ref{fig.RBH}. Additionally, we plot the geodesics of the point particle obtained independently of the SPH code with the same initial conditions. Based on \cite{Banerjee2023}, the SPH core center-of-mass orbits deviate from the test particle geodesic when the mass ratio ($q = M/M_\star$) is smaller than $10^3$. For our case, the mass ratio is approximately $3.3 \times 10^4$, explaining the lesser deviations in the trajectories in our work.

\begin{figure*}
	\centering 
	\includegraphics[scale=0.25]{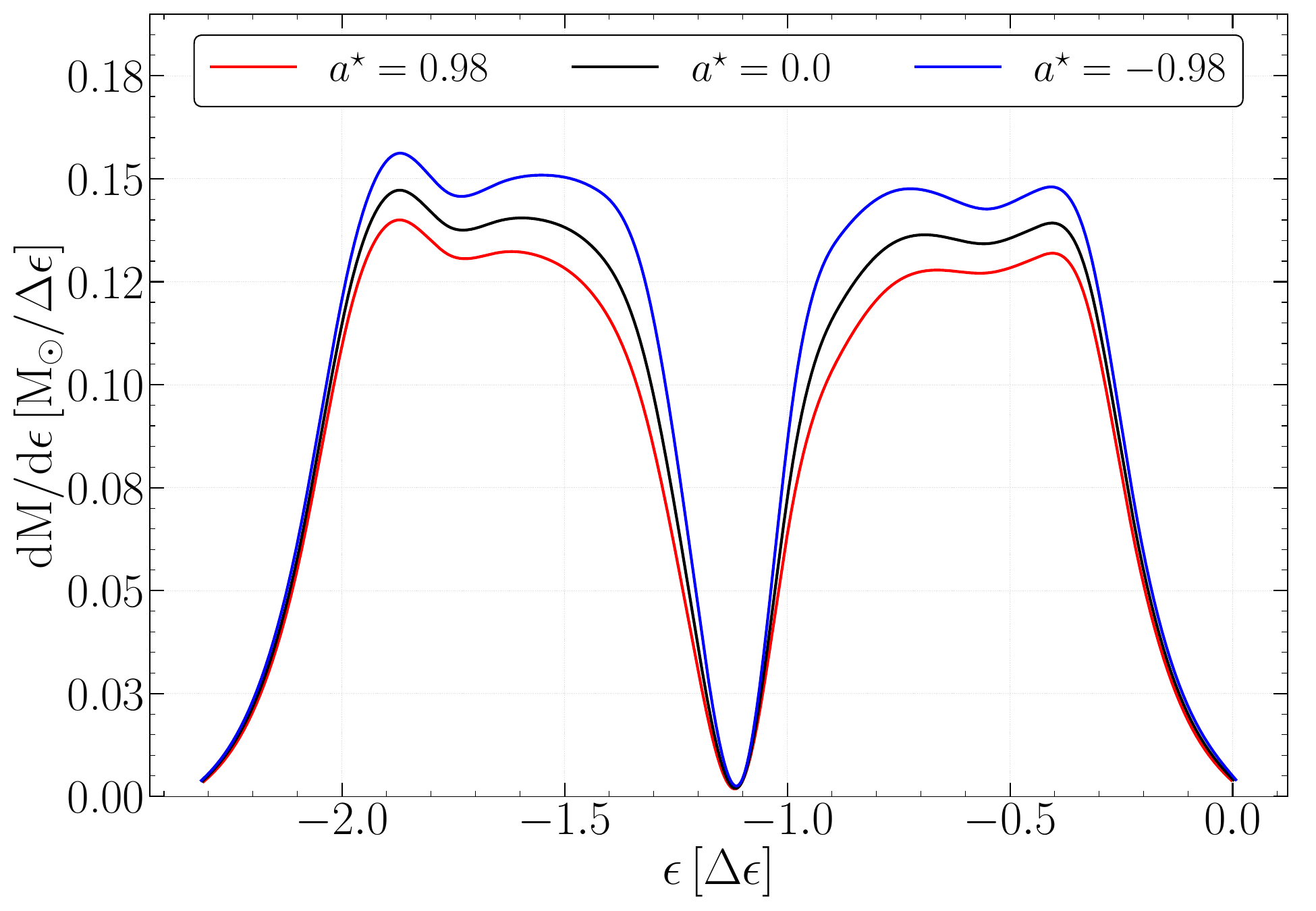}
	\includegraphics[scale=0.25]{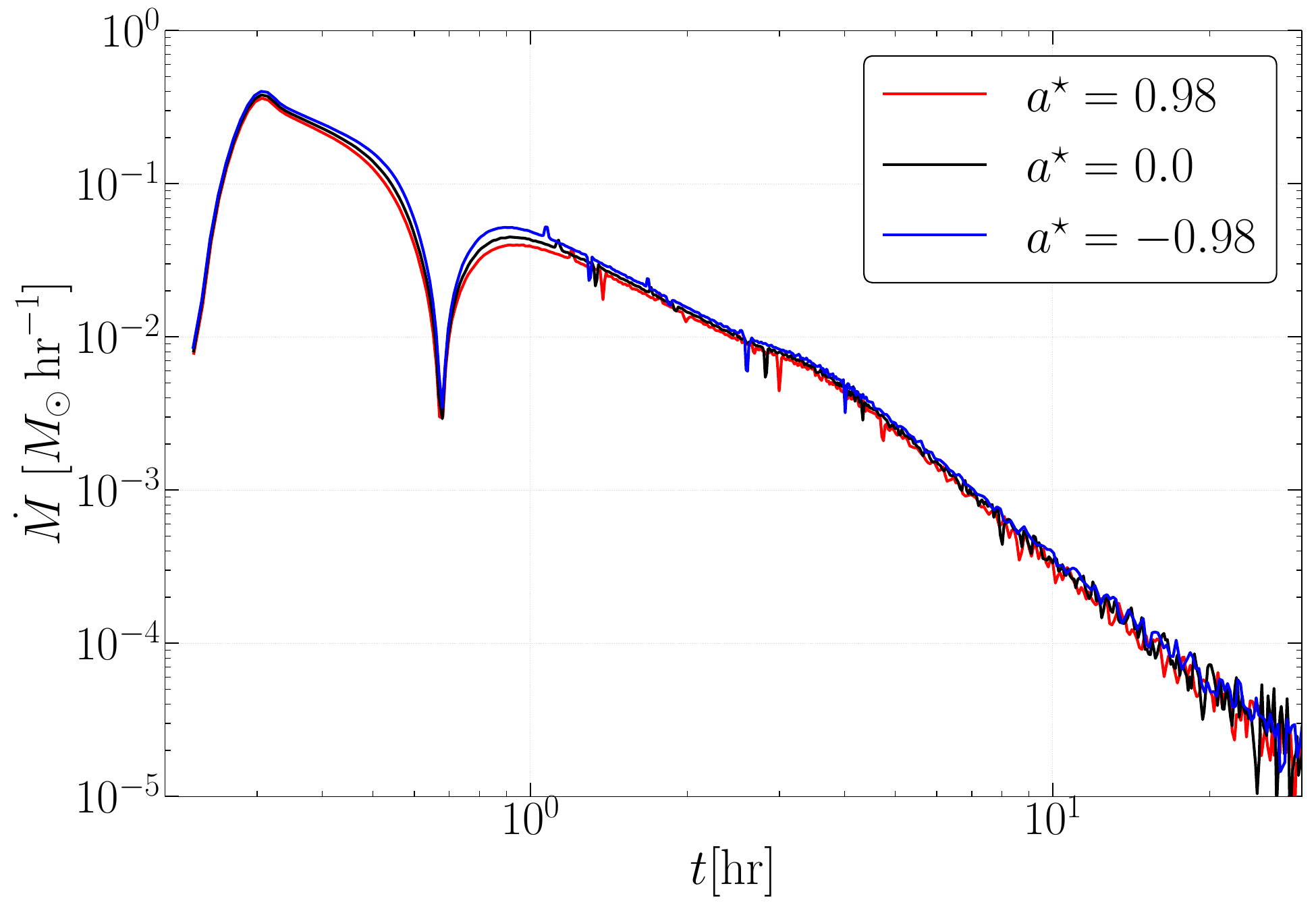}		
	\caption{\textbf{Left Panel:} Variation of debris differential mass distribution with specific energy. \textbf{Right Panel:} Fallback rates are plotted against time in hours. The different values of the BH spin parameter, $a^\star$, are mentioned in the legends.}
	\label{fig.RBH1}
\end{figure*}

Tidal disruption in elliptical trajectories can significantly differ from parabolic ones. If the eccentricity of the stellar object's orbit is $e < e_{\rm crit}^-$, no unbound debris is produced. In contrast, if $e > e_{\rm crit}^+$, no bound debris is produced. Here, $e_{\rm crit}^\pm = 1 \pm (2/\beta)q^{-1/3}$, with $q$ representing the mass ratio ($q = M/M_\star$), as discussed in \cite{Hayasaki2018}. For our simulation parameters, the eccentricity falls below $e_{\rm crit}^-$, suggesting that all debris should theoretically be bound to the black hole. To confirm this, we compute the debris differential mass distribution ($\rm{dM/d\epsilon}$) with specific energy ($\epsilon$). In the left panel of Figure \ref{fig.RBH1}, we present $\rm{dM/d\epsilon}$ as a function of $\epsilon$ for various values of $a^\star$. The specific energy is normalized by the spread in specific energy of the stellar debris, denoted by $\Delta\epsilon = G M R_{\rm wd}/r_t^2$. Post-disruption snapshots at $t \approx 0.13\,\rm{hr}$ were used to calculate $\rm{dM/d\epsilon}$ for different values of $a^\star$, with core particles being removed from the calculation, as we were only interested in the debris.

From the left panel of Figure \ref{fig.RBH1}, it is evident that almost all debris is bound to the BH for all $a^\star$ values, with the mass of the bound debris being the initial WD mass minus the core mass. The bound debris, having negative specific energy, is accreted by the BH. 
On the other hand, the surviving core, predicted to follow an elliptical trajectory, is expected to return to the pericenter, where it may lose additional mass in consecutive interactions.
However, in this work, we focus solely on the single interaction between the WD and BH, so we stop the simulation after the first pericenter passage once all bound debris is accreted by the BH. Multiple passages could result in interesting physics, which we leave for a future study. To calculate the fallback rate of the bound debris, we tracked the amount of mass falling onto the black hole through the accretion radius ($3\,r_t \simeq 82\,r_g$), then took the numerical derivative of the data to obtain $\dot{M}$, shown in the right panel of Figure \ref{fig.RBH1} for different values of $a^\star$. An interesting feature of the fallback rate is the presence of two peaks. The left arm in the left panel of Figure \ref{fig.RBH1} is accreted first by the black hole, resulting in the first peak in the fallback rate plots in the right panel of the same figure.

Following this, the right arm is accreted by the black hole, leading to the second peak in the fallback rate plot. As more mass is disrupted for $a^\star = -0.98$ compared to $a^\star = 0.98$, the fallback rate plot for $a^\star = -0.98$ is slightly higher than for $a^\star = 0.98$, although not significantly. For all $a^\star$ values, the first peak is approximately one order of magnitude higher than the second peak. 
We calculate the ratio of left arm mass to right arm mass to be approximately 1.1, which contributes to the higher value of the first peak compared to the second peak in the fallback rate plots. However, this mass difference alone does not account for the order of magnitude difference between the peaks. The first peak's higher fallback rate is primarily due to the faster return rate of the more tightly bound debris associated with the first peak, which has more negative specific energy, leading to a quicker fallback.
While the most bound debris fallback times for $a^\star = -0.98$ and $a^\star = 0.98$ are almost identical, our numerical results indicate that the most bound debris fallback time for $a^\star = -0.98$ is slightly earlier than for $a^\star = 0.98$.
The late-time fallback rate for all $a^\star$ values is similar, making it challenging to distinguish the black hole's spin from the fallback rate plots.

Next, we study the impact of the initial uniform rotation of the WD on the bound debris and fallback rates in the presence of a non-spinning BH. To ensure that the initial spin of the WD does not significantly distort its spherical shape, we choose a maximum spin period that maintains an approximately same tidal radius for all WDs with different spin values ($\lambda$). This way, all WDs experience the same impact parameter at the pericenter distance. In this scenario, initial stellar spin becomes the key factor in determining the fate of the debris, which we will discuss shortly. For this study, we keep the orbital angular momentum along the positive $z$ direction, and the spin angular momentum is always along the orbital angular momentum direction. When the spin angular momentum is parallel to the orbital angular momentum, it is termed prograde rotation of the WD, with $\lambda > 0$. When the spin angular momentum is anti-parallel to the orbital angular momentum, it is termed retrograde rotation of the WD, with $\lambda < 0$.

\begin{figure*}
	\centering 
	\includegraphics[scale=0.25]{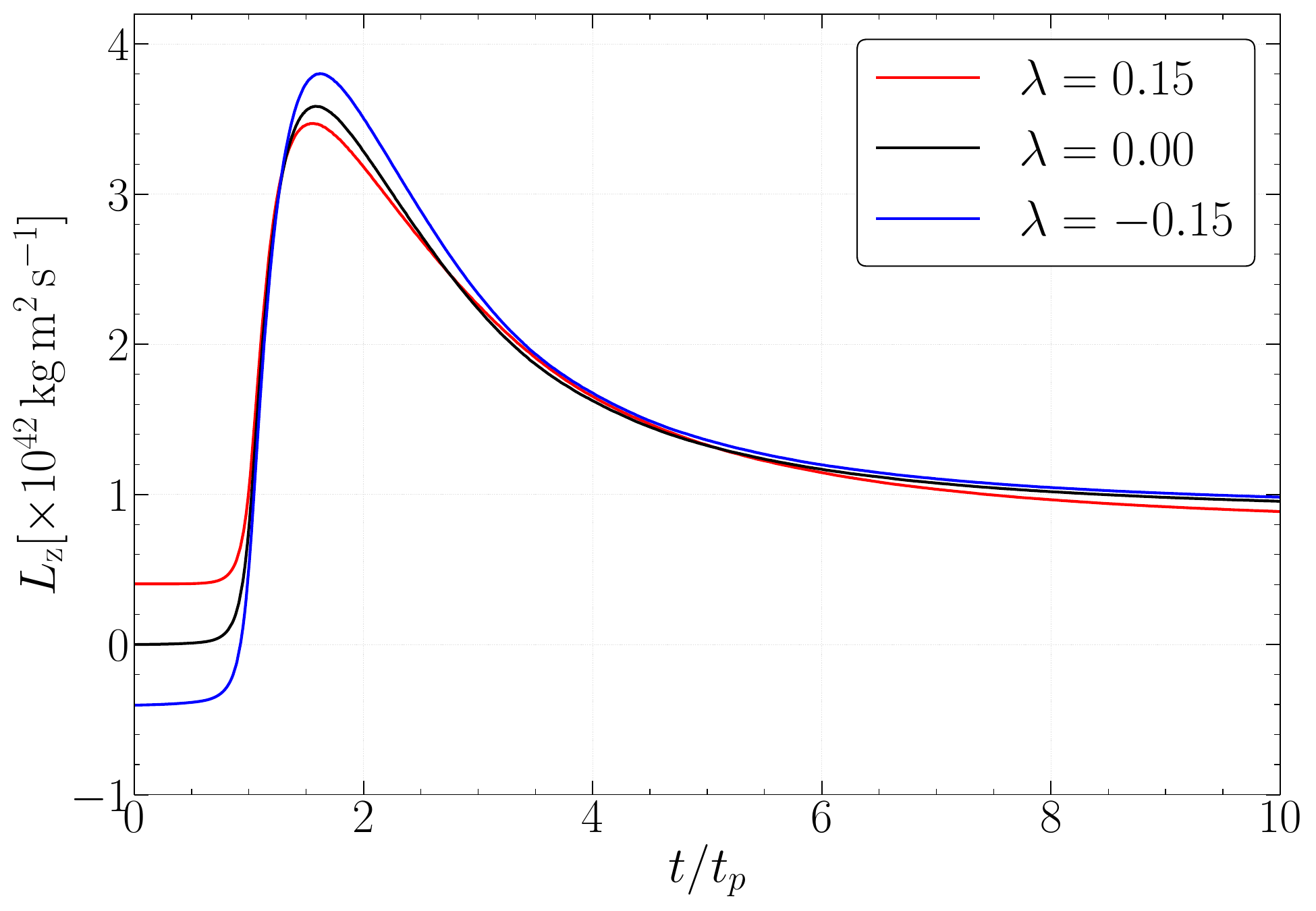}
	\includegraphics[scale=0.25]{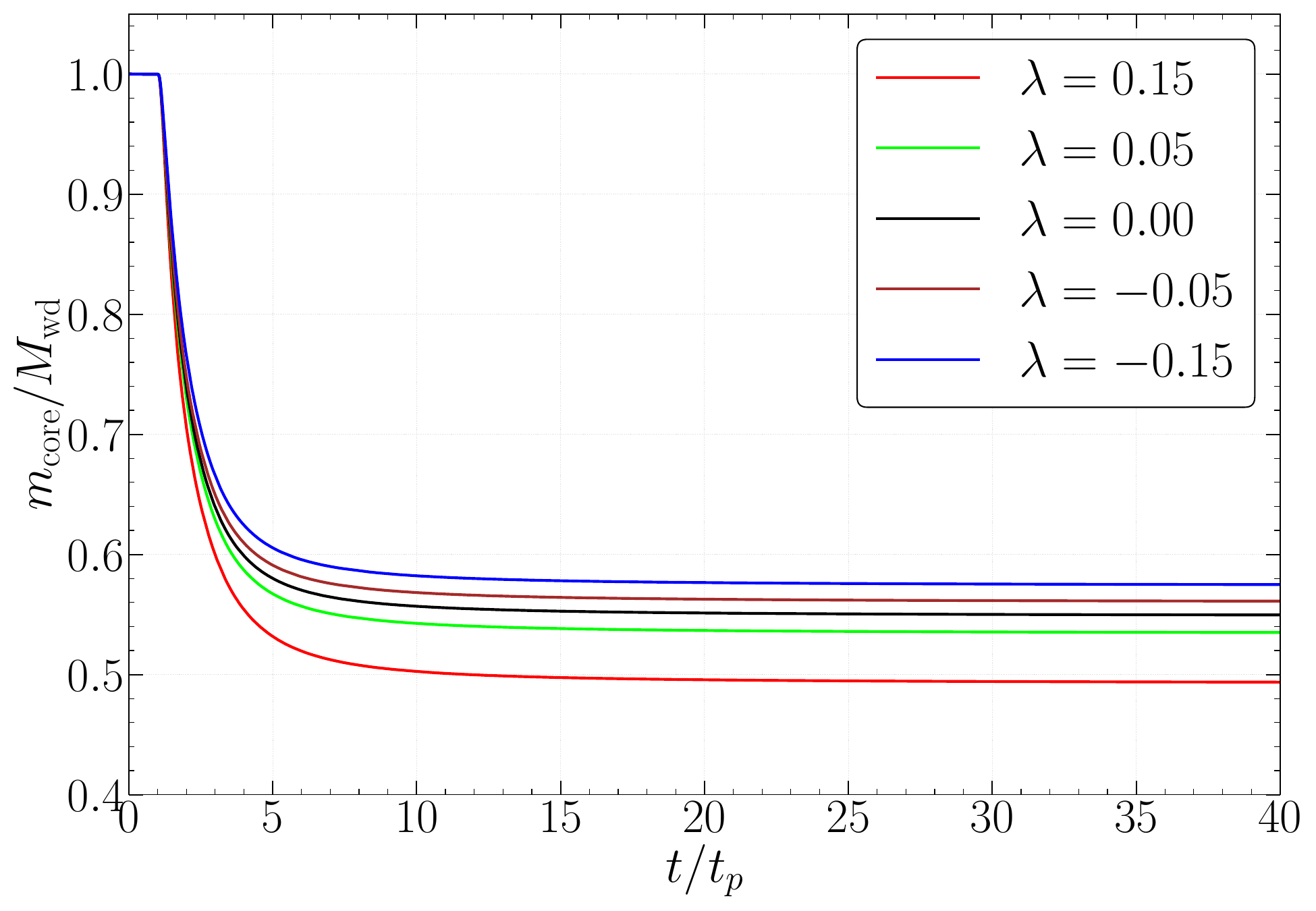}		
	\caption{\textbf{Left Panel:} Variation of the $z$-component of the spin angular momentum with time for bound particles. \textbf{Right Panel:} Variation of the core mass fraction, $m_{\rm core}/M_{\rm wd}$, plotted against normalized time. The different values of the break-up fractions $\lambda$ for different spinning WDs are mentioned in the legend.}
	\label{fig.RWD1}
\end{figure*}

\begin{figure*}
	\centering 
	\includegraphics[scale=0.25]{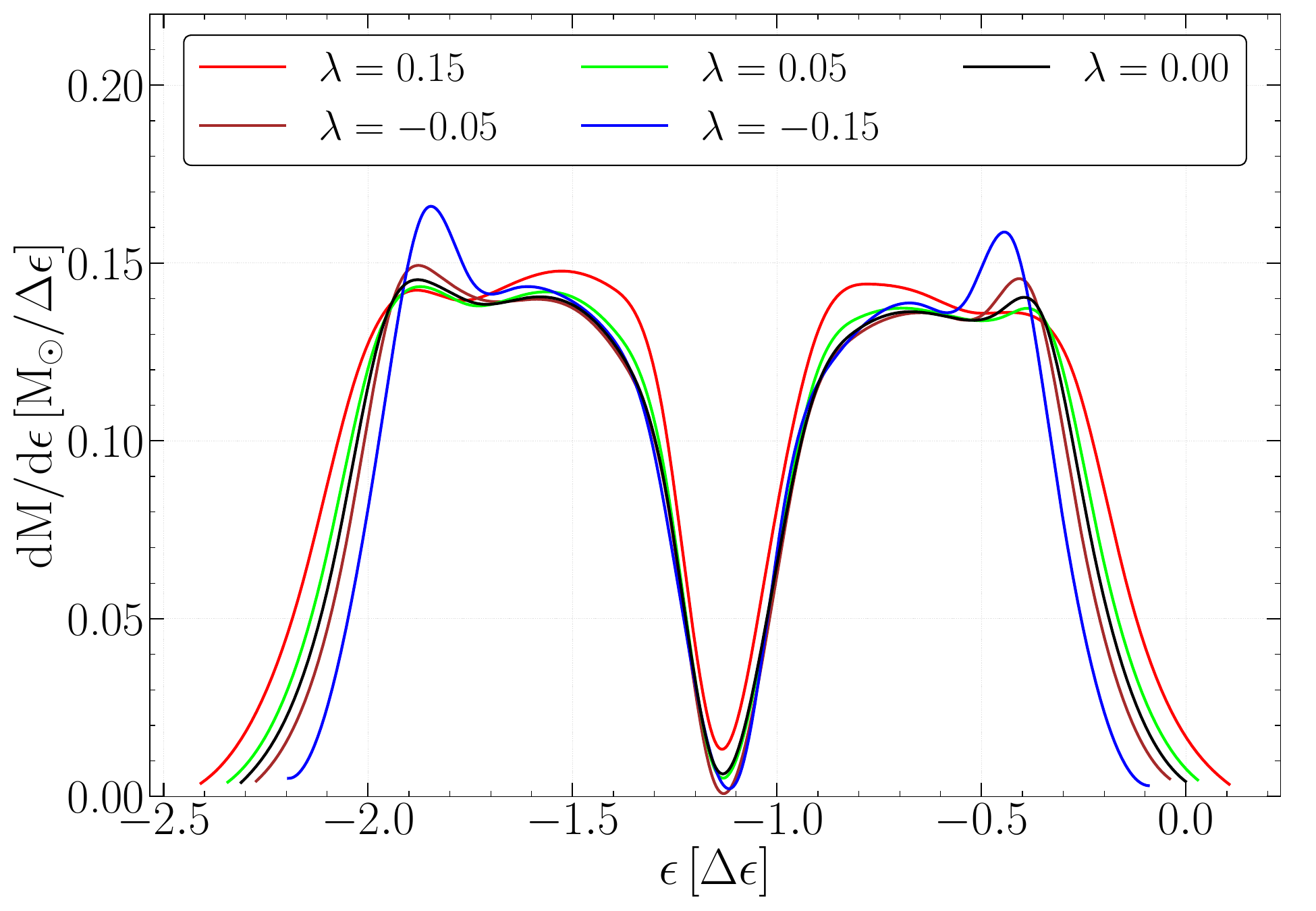}
	\includegraphics[scale=0.25]{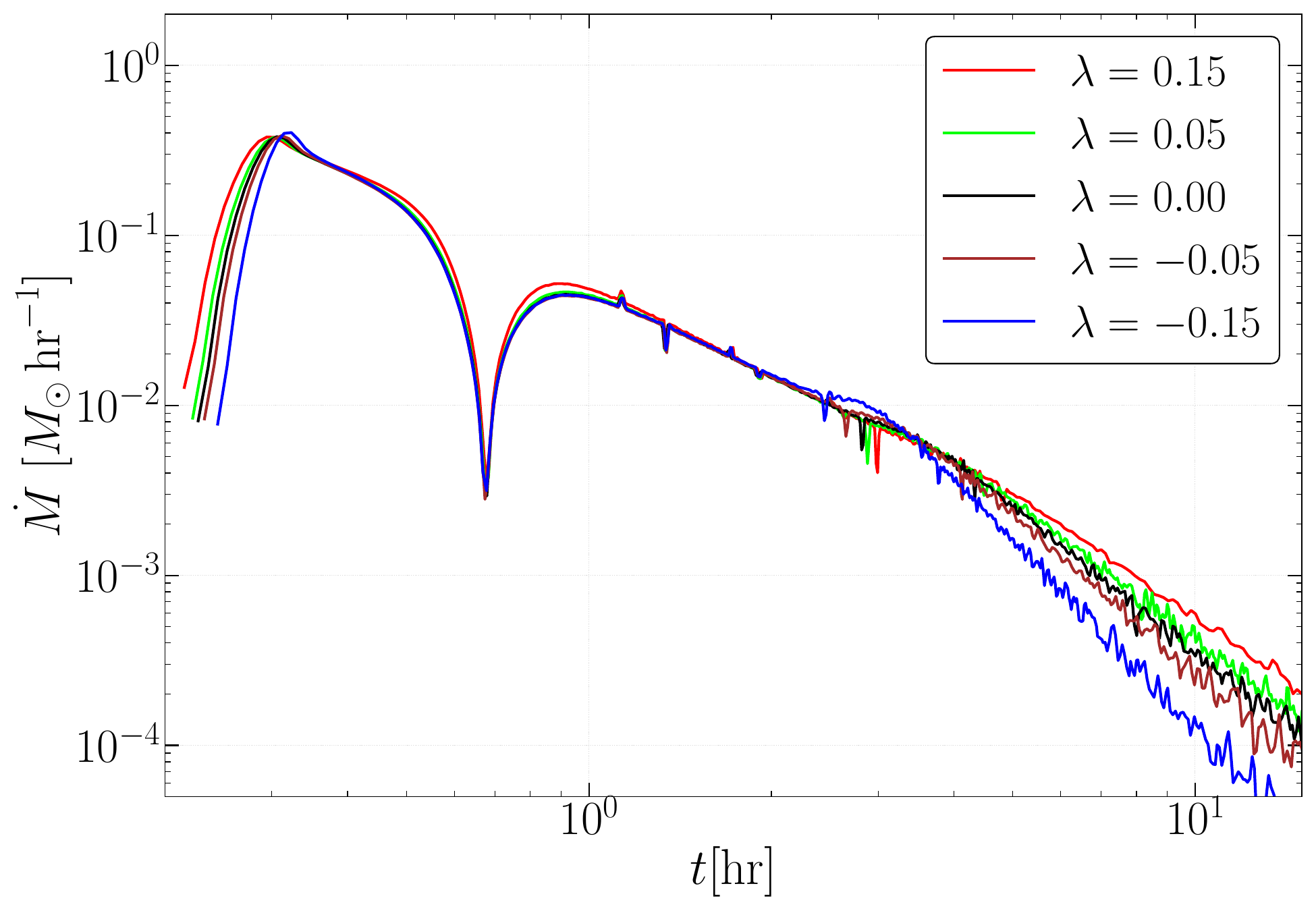}		
	\caption{\textbf{Left Panel:}Variation of debris differential mass distribution with specific energy. \textbf{Right Panel:} Fallback rates are plotted against time in hours. The different values of the break-up fractions ($\lambda$) for different spinning WDs are mentioned in the legend.}
	\label{fig.RWD}
\end{figure*}

As the WD approaches the BH, differential gravitational forces from the BH start to exert tidal deformation on the WD. This deformation results in stretching along the direction towards the BH and compression in the orthogonal directions of the WD even before reaching the pericenter distance $r_p$. When the WD reaches $r_p$, the tidal force becomes significant, leading to a net torque exerted on the WD. This torque induces prograde rotation with respect to the orbital motion of the WD. We refer the reader to Section 2.2 of \cite{Sacchi2019}, where similar issues are discussed.

To extend the above discussion further, we quantify the rotation of the remnant core by presenting the variation in the $z$-component of the spin angular momentum $L_z$ of the core particles as a function of normalized time in the left panel of Figure \ref{fig.RWD1}, considering three values of $\lambda$. Since the orbital motion lies in the $x$–$y$ plane, $L_z$ captures the spin imparted by tidal torques during the encounter. As the WD approaches pericentre, the strong tidal field induces a non-axisymmetric deformation whose quadrupole misalignment with the black hole's gravitational gradient leads to a net torque, transferring orbital angular momentum into stellar spin. This effect, largely independent of the initial spin orientation, causes $L_z$ to increase sharply near pericentre and saturate afterward as the tidal interaction weakens. Notably, in the retrograde case ($\lambda = -0.15$), this torque is strong enough to reverse the spin direction. While the peak values differ slightly due to varying bound masses post-stripping, the final $L_z$ converges across all $\lambda$, indicating that the initial spin is effectively erased after a single passage.

Continuing the analysis of how the direction of 
rotation affects mass disruption, we show the variation of the core mass fraction over normalized time for different values of $\lambda$ in the right panel of Figure \ref{fig.RWD1}. It is evident that the initial WD rotation either helps or hinders the disruption, as tidal forces also impart a rotation. The initial prograde rotation of the WD is in the same direction as the rotation induced by the tidal interaction, resulting in more mass being disrupted and a lower core mass fraction as we go from $\lambda = 0$ to $\lambda > 0$. In contrast, the initial retrograde rotation of the WD opposes the rotation induced by the tidal interaction, causing the black hole to change the spin direction (evident from the left panel of the same figure), resulting in less mass disruption and a higher core mass fraction as we go from 
$\lambda = 0$ to $\lambda < 0$.

The analytical calculation by \cite{Sacchi2019} shows that the spread in specific energy after tidal interaction varies with $\lambda$ (called $\alpha$ in their work) when the rotation axis is perpendicular to the orbital plane, as in our study (though they use a Newtonian potential). The energy spread increases for $\lambda > 0$ and decreases for $\lambda < 0$. We observe the same behavior in our results. In the left panel of Figure \ref{fig.RWD}, we present the differential mass distribution of the debris as a function of specific energy for different values of the break-up fraction. The snapshots used for this calculation are around $t \approx 0.13\,\rm{hr}$ before mass accretion occurs. For $\lambda > 0$, the profile of $\rm{dM/d\epsilon}$ is more spread out compared to $\lambda < 0$. The prograde rotation of the WD, with a larger spread in $\epsilon$, leads to more disruption and less mass in the core, as seen in the right panel of Figure \ref{fig.RWD1}. 

\begin{figure*}
	\centering 
	\includegraphics[scale=0.25]{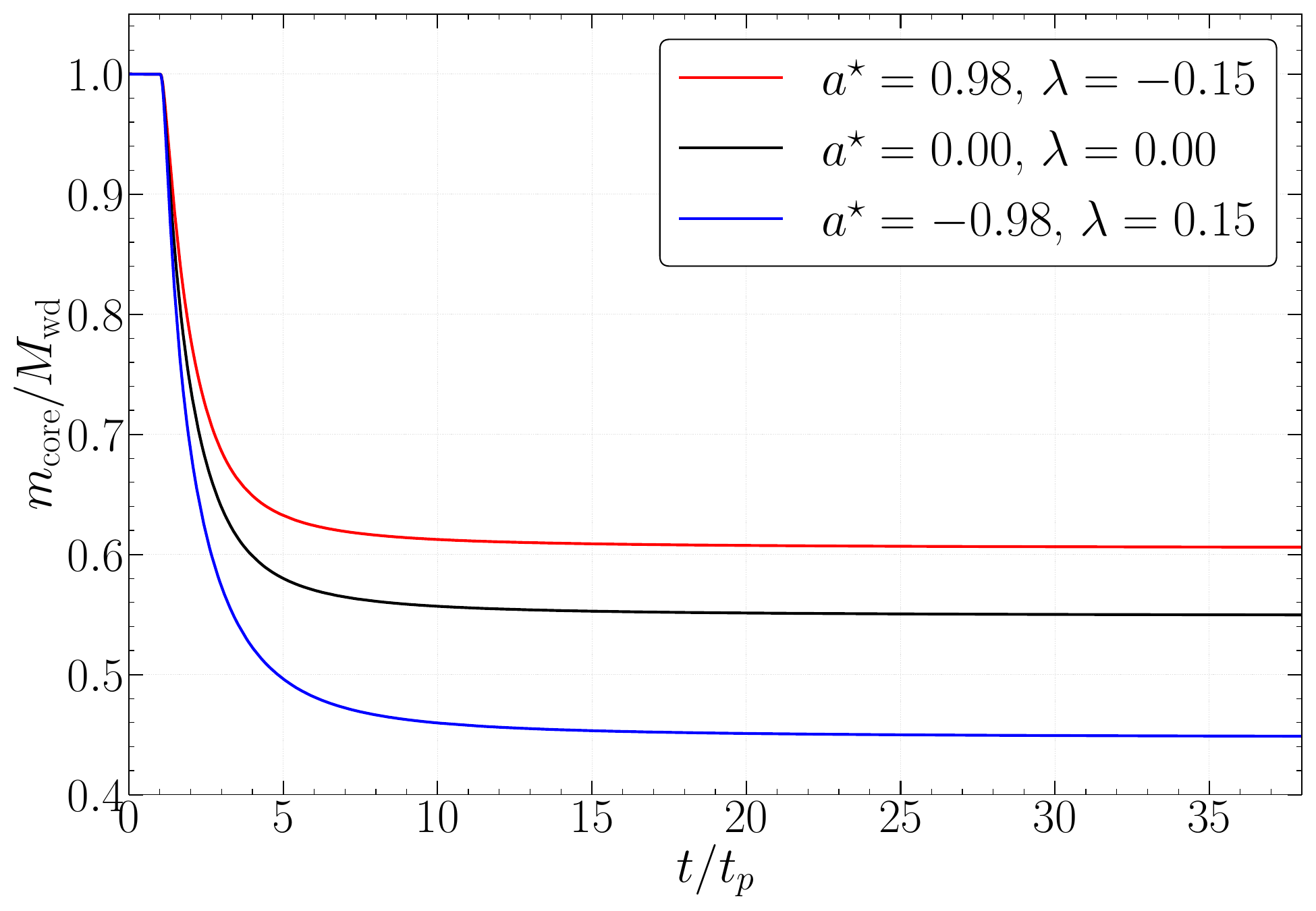}
	\includegraphics[scale=0.25]{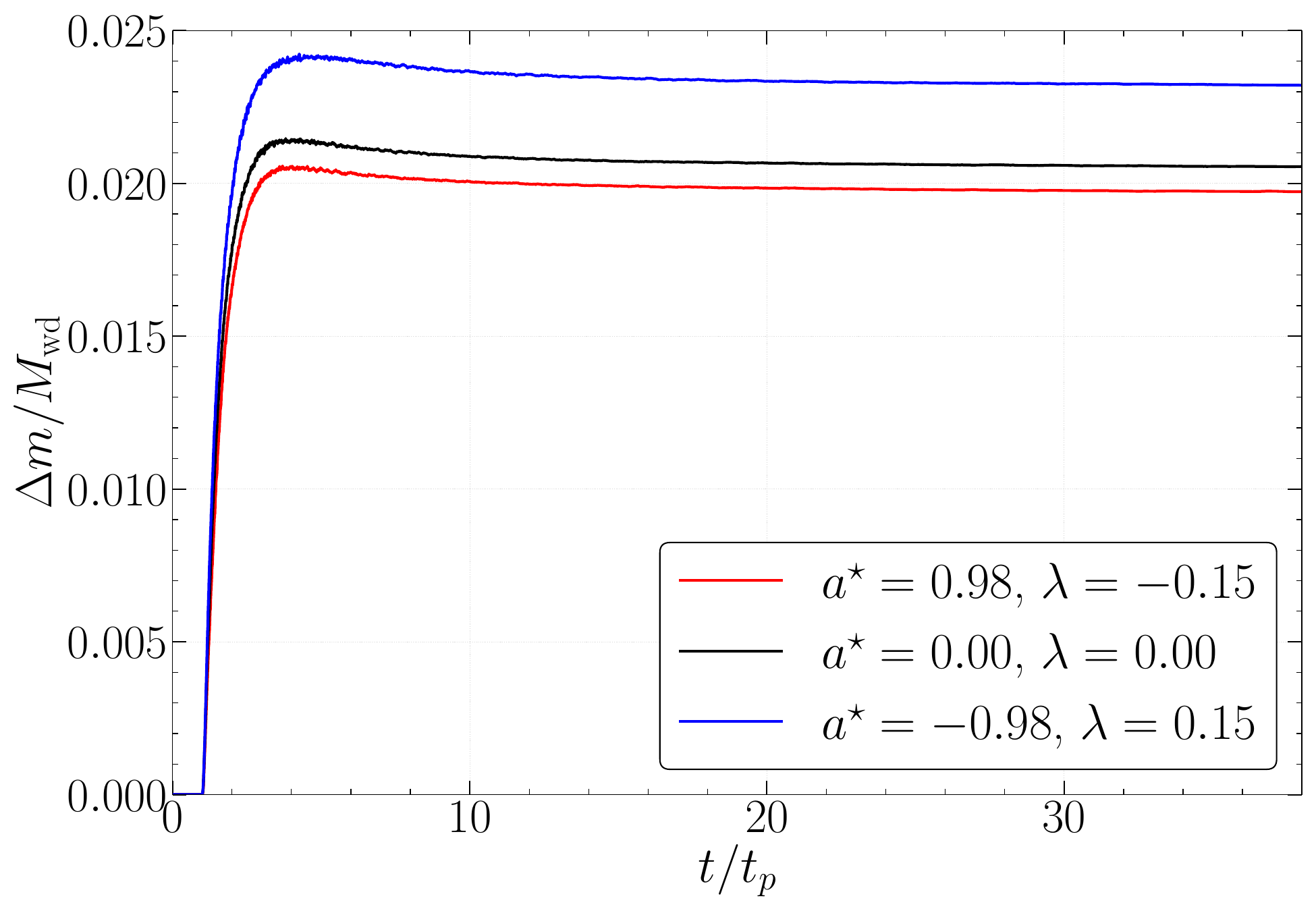}		
	\caption{\textbf{Left Panel:} Core mass fraction variation, $m_{\rm core}/M_{\rm wd}$, plotted against normalized time. \textbf{Right Panel:} Mass difference between the two tidal tails relative to the initial WD mass, plotted against normalized time. Legends indicate the values of BH spin parameters ($a^\star$) and break-up fraction of the WDs ($\lambda$).}
	\label{fig.RBHRWD}
\end{figure*}

\begin{figure*}
	\centering 
	\includegraphics[scale=0.25]{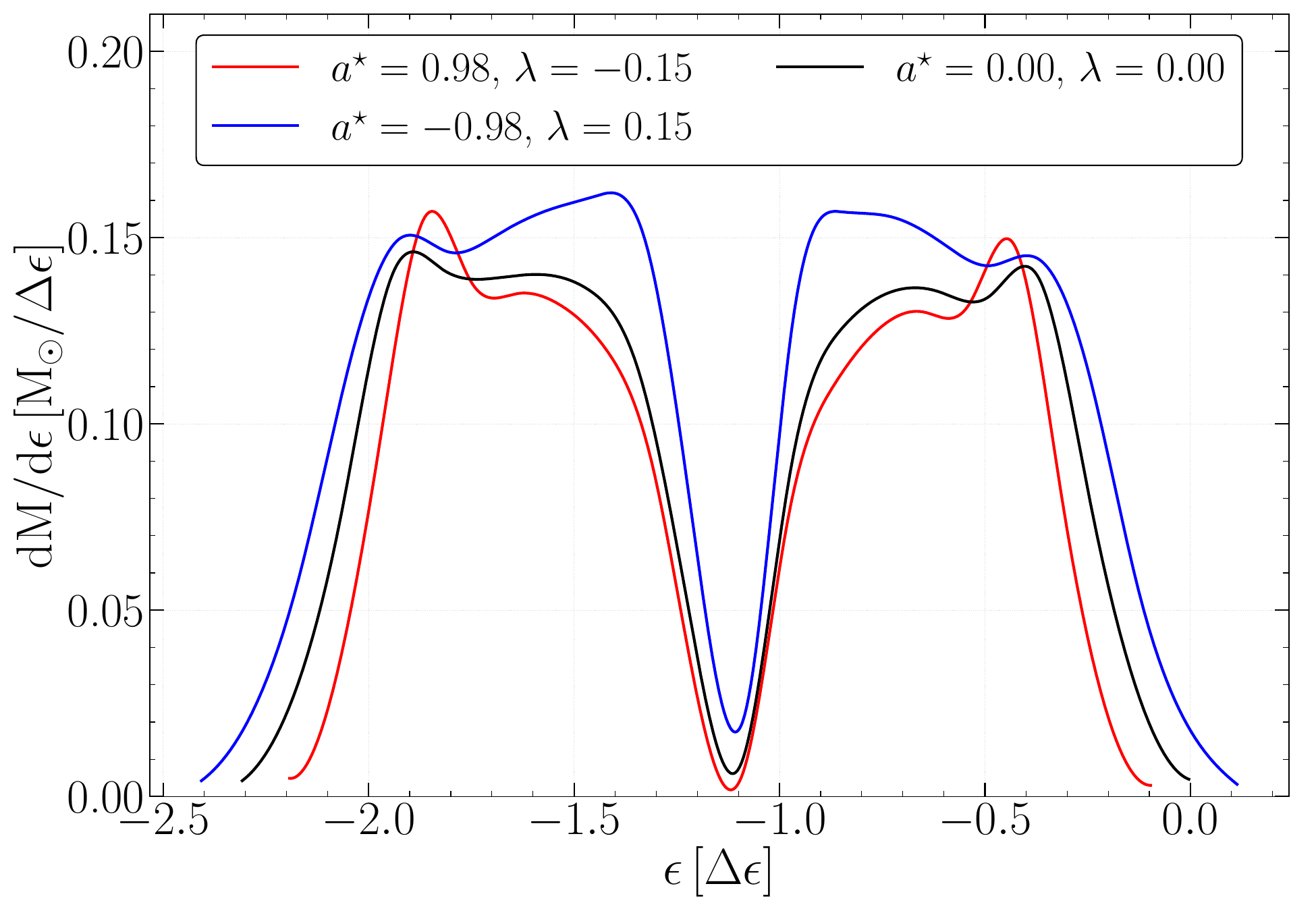}
	\includegraphics[scale=0.25]{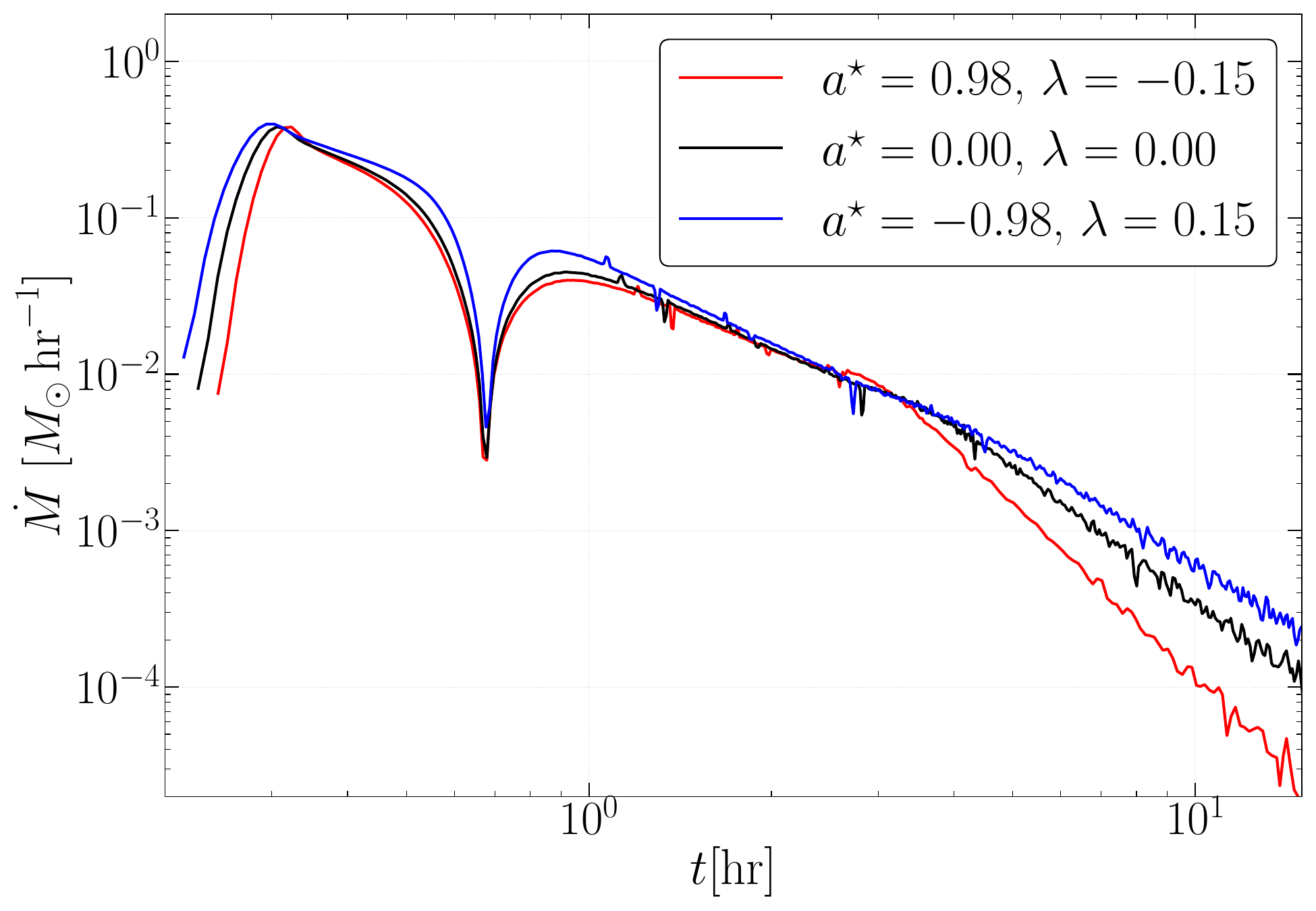}		
	\caption{\textbf{Left Panel:} Variation of debris differential mass distribution with specific energy. \textbf{Right Panel:} Fallback rates are plotted against time in hours. Legends indicate the values of BH spin parameters ($a^\star$) and break-up fraction of the WDs ($\lambda$).}
	\label{fig.RBHRWD1}
\end{figure*}

In contrast, the retrograde rotation, with a smaller spread in $\epsilon$, is more resistant to disruption, resulting in more mass in the core. In the right panel of Figure \ref{fig.RWD}, we plot the fallback rate over time in hours for different values of $\lambda$. The fallback rate plots exhibit similar features to those in the right panel of Figure \ref{fig.RBH1}: two peaks and a dip, with the first peak being higher than the second. After disruption, the debris bound to the black hole is less compact with prograde spin compared to retrograde spin. Consequently, the most bound debris falls back onto the black hole more quickly in the prograde case because it is more dispersed. We calculate that the most bound debris falls back approximately $14\%$ earlier for a prograde spin compared to a retrograde spin of the WD. We also find that the late-time fallback rate varies with $\lambda$. This variation in the late-time fallback rate is not present in the right panel of Figure \ref{fig.RBH1}.
This could be significant from an observational perspective, particularly in scenarios where multiple disruptions occur during each pericenter passage, potentially contributing to the late-time fallback rate. Additionally, it is important to note that in those scenarios, the actual late-time fallback could be influenced by these subsequent disruptions, making it an interesting area for future studies to explore how the stripped material from the second and subsequent interactions could add to the existing debris to alter the fallback rates. A detailed analysis of the exact variation in the late-time power law, considering these factors, is a topic worth studying in future.

Next, we explore the combined rotation effects of the BH and WD on the material bound to the BH and the resulting fallback rates. In the presence of a spinning BH and a non-spinning WD, we observe that for retrograde spin ($a^\star < 0$), there is more mass disruption from the initial WD compared to the prograde spin of the BH ($a^\star > 0$). Conversely, in the presence of a non-spinning BH and a spinning WD, the prograde spin of the WD (parallel to the orbital direction, $\lambda > 0$) leads to more disruption from the initial WD than the retrograde spin (anti-parallel to the orbital direction, $\lambda < 0$) of the initial WD. Therefore, to achieve maximum disruption from the initial WD, coupling the retrograde spin of the BH with the prograde spin of the WD is required, while for minimum disruption, coupling the prograde spin of the BH with the retrograde spin of the WD is necessary. All other combinations fall between these chosen configurations.

In the left panel of Figure \ref{fig.RBHRWD}, we display the variation in core mass fraction over normalized time. Notably, we observe a nearly $35\%$ relative change in core mass fraction from the minimum to the maximum disruption, contrasting with around $14\%$ for the scenario with a spinning BH and non-spinning WD, and approximately $16\%$ for a non-spinning BH and spinning WD 
\footnote{We want to remind the reader that when we mention the `relative change' of a quantity $A_2$ with respect to $A_1$, we are referring to $|A_2-A_1|/A_1$. This convention is followed consistently throughout the article.}. 

Hence, the ``coupled'' rotation significantly influences the mass disruption from the initial WD during tidal interaction, which eventually reflects in the core mass fraction variation with time.
The increased precession of the retrograde orbit of the BH, combined with the initial spin of the WD in the direction of the WD's orbital motion, enhances the disruption from the initial WD. 
In contrast, a prograde BH spin combined with a retrograde spin of the WD makes the WD more resilient to tidal disruption. We show the variation in the mass difference between the two tidal tails ($\Delta m$) over normalized time in the right panel of Figure \ref{fig.RBHRWD}. Here, $\Delta m$ is normalized with respect to the initial WD mass. Specifically, $\Delta m = m_1 - m_2$, where $m_1$ is the mass of the tidal tail closer to the BH and $m_2$ is the mass of the tidal tail farther from the BH.
The individual evolutions of $m_1$ and $m_2$ are provided in Appendix~\ref{app:m1m2}.
The mass difference increases notably when the BH is in a retrograde orbit and the WD has a prograde spin. In partial TDEs, the asymmetric mass loss in the two tidal tails can lead to interesting phenomena. Previous studies by \cite{Manukian2013} and \cite{Gafton2015} have shown that asymmetric mass loss results in a ``kick" of the surviving core in the context of TDEs by super-massive BHs when the incoming trajectory is parabolic.  
However, the final fate of the core, whether it remains bound or becomes unbound, also depends on other factors, such as the injection of orbital energy into the oscillation modes of the stellar object through tides. As discussed in \cite{Broggi2024}, this energy transfer can play a significant role beyond a certain pericenter distance for different stellar models. If the energy dissipated from the orbital energy into oscillation modes surpasses the orbital energy gain from the asymmetric tidal tail formation, the core could remain tightly bound to the black hole. Given this, we compute the orbital energy change of the self-bound core for different ``coupled'' rotation combinations.
The relative change in the specific orbital energy of the core compared to the initial specific orbital energy of the WD is about $10^{-3}\,\%$ for the maximum mass disruption combination and about $0.6 \times 10^{-3}\,\%$ for the minimum mass disruption combination. However, this increase in specific orbital energy is not enough for the disrupted WD to escape the BH's influence. Therefore, it may require multiple interactions to completely eject the WD from the BH's influence. Such multiple interactions involving an IMBH and a main sequence star have recently been reported in \cite{Kiroglu}.

To analyze the fallback of the bound debris, we first present the differential mass distribution of the debris with specific energy in the left panel of Figure \ref{fig.RBHRWD1} for three different combinations of spin as mentioned in the legend. The snapshots used to compute the mass distribution for these combinations are taken around $t \approx 0.13\,\rm{hr}$. The spread in the specific orbital energy of the debris is mainly due to the rotation of the WD, with only a slight dependence on the BH rotation, as seen in the left panels of Figure \ref{fig.RBH1} and Figure \ref{fig.RWD}. However, the mass in the two arms is influenced by both the BH spin and the WD spin. In the right panel of Figure \ref{fig.RBHRWD1}, we show the mass fallback rate as a function of time in hours. We notice a slight ``broadening'' of the fallback curve corresponding to the
maximum mass disruption compared to the ones in Figures \ref{fig.RBH1} and \ref{fig.RWD}. Also, the greater mass in the left arm compared to the right arm results in a higher first peak than the second peak in the fallback rate plot. The most bound debris fallback time varies among different combinations, with a relative change of about $15\%$ from the coupling of a prograde BH and retrograde WD to the coupling of a retrograde BH and prograde WD. 

However, as previously discussed, the initial spin of the WD has a more significant impact on the most bound debris fallback time than the rotation of the BH. Therefore, in the case of ``coupled" rotation, the change in the most bound debris fallback rate time is mainly due to the spin of the WD. We observe a dispersion in the fallback rates for the different spin combinations during the late part of the first peak, around $t \approx 0.5 - 0.6 \,\rm{hr}$, which does not appear for individual spins. The second peak in the fallback rate plots for different combinations also shows differences due to the coupled rotation effect, while individual spins do not significantly influence the second peak values. Finally, we observe that the late-time fallback rates vary for different combinations. However, this variation is mainly due to the WD rotation, as the late-time fallback rate remains almost the same regardless of the BH rotation, for non-spinning WDs.

\subsection{Gravitational Wave Emission}

Once the WD interacts with the BH near $r_p$, it emits gravitational
waves (GWs) due to the variation of the mass quadrupole of the star-BH system, see \cite{kob2004}. Following the disruption, the stretched debris is less compact to emit robust gravitational waves after leaving the closest approach. As a result, the
detectable gravitational wave signal exhibits burst-like behavior, characterized by an amplitude $h$, signal duration $\tau$, and frequency $f \sim 1/\tau$ \citep{Toscani2022}.

\begin{figure*}
	\centering
	\includegraphics[scale=0.4]{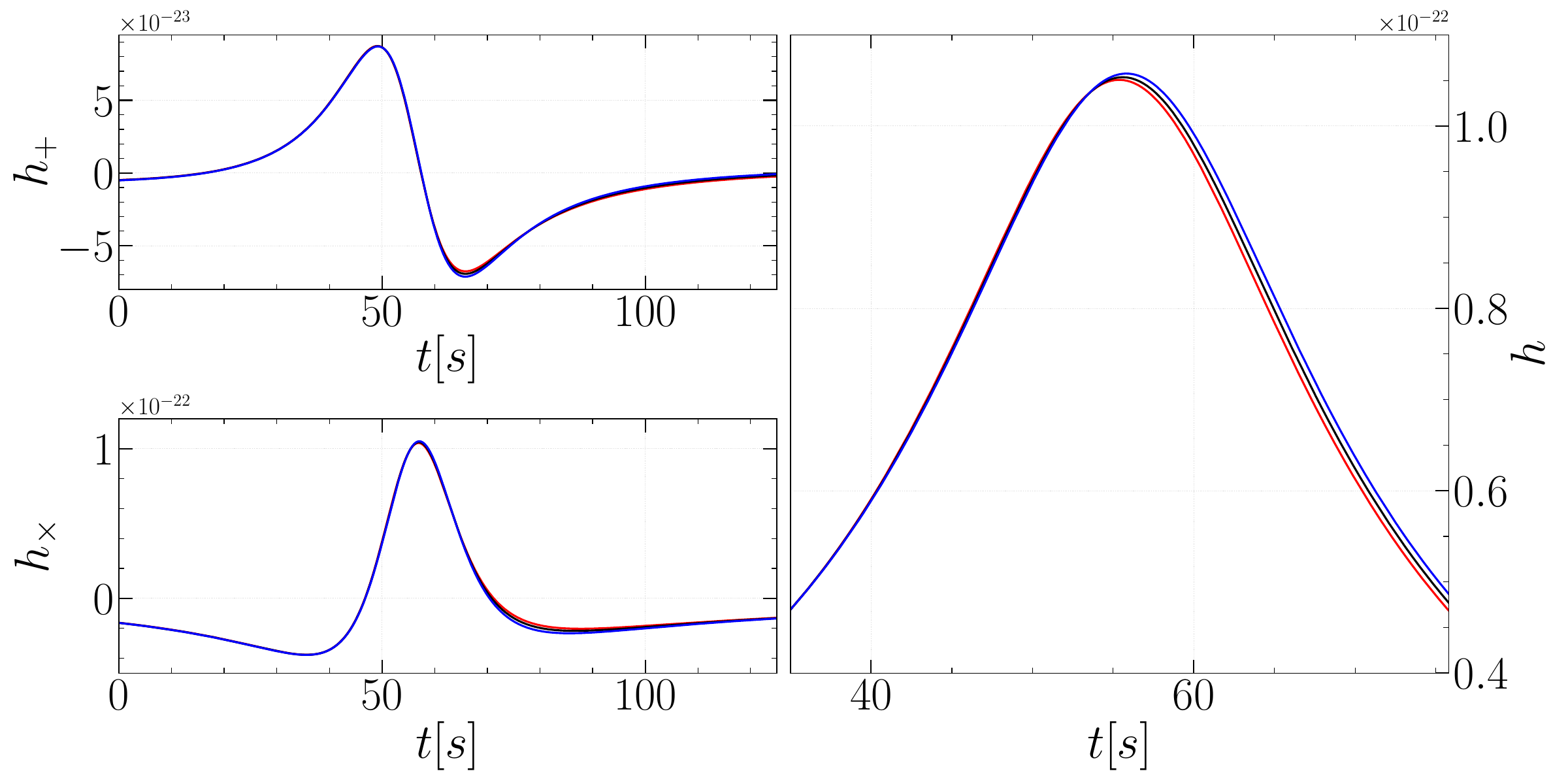}
	\includegraphics[scale=0.6]{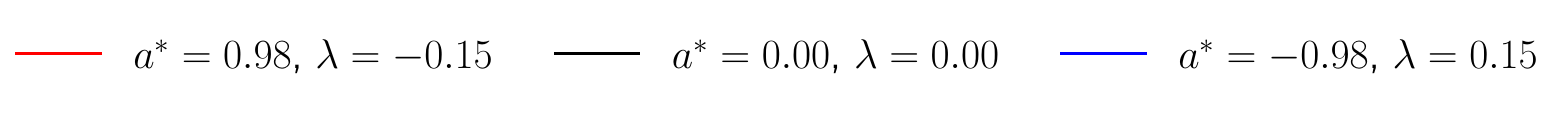}
	\caption{GW emission resulting from the tidal interaction between a WD with a mass of $0.3 M_\odot$ and an IMBH with a mass of $10^4 M_\odot$. The source is located at a distance of $20$ Mpc. \textbf{Top left Panel:} Variation of + polarization with time in seconds. \textbf{Bottom left Panel:} Variation of $\times$ polarization with time in seconds. \textbf{Right Panel:} Variation of the root-square-sum amplitude with time in seconds. Different curves correspond to various values of $a^\star$ and $\lambda$ as indicated in the legend.}
	\label{fig.gw}
\end{figure*}

To calculate GW amplitudes numerically within the SPH framework, we adopted the methodology outlined in \cite{Rosswog2009}. Using the second-time derivative of the reduced quadrupole moment, we calculate the polarization amplitudes, $h_+(t)$, $h_\times(t)$, and the root-square-sum amplitude, $\lvert h(t) \rvert= \sqrt{\lvert h_+(t) \rvert ^2 + \lvert h_\times(t) \rvert ^2}$ over time for different tidal disruption scenarios. 
Figure~\ref{fig.gw} presents the gravitational waveforms $h_+(t)$, $h_\times(t)$, and $\lvert h(t) \rvert$ for three representative combinations of BH and WD spins, as labeled in the legend.

As we change the combination from $a^\star = -0.98,\, \lambda = 0.15$ to $a^\star = 0.98,\,\lambda=-0.15$, the peak of the root-sum-square amplitude $h$ shifts to earlier times and decreases. The relative change in the peak of $h_{\rm max}$ between these two extreme combinations is approximately $0.7\%$. To understand the reason behind the difference, we further studied the GW amplitudes considering BH and WD rotations individually. We find all GW strains are similar and do not deviate from each other due to the tidal interaction between non-spinning BHs and spinning WDs. In contrast, in TDEs incorporating spinning BHs and non-spinning WDs, deviations arise for different BH spin parameters. It is worth mentioning that \cite{Toscani2022} also observed a similar behavior of the root-sum-square amplitude with BH spin (see their Figure 7), noting a significant variation in the peak value of $h$ for their chosen parameters as they consider higher $\beta$ values for tidal interaction. Hence, we conclude that the difference in the GW strains for different ``coupled'' combinations is mainly contributed by the BH spin.

For our simulation parameters, where $a^\star = 0$ and $\lambda = 0$, we calculate the amplitude and frequency of the gravitational wave signal using the analytical expressions from Equations 12 and 13 in \cite{Toscani2022}. We found the amplitude to be approximately $0.73 \times 10^{-22}$ and the frequency to be around $10^{-2}$ Hz. Assuming a source to observer distance of $d = 20 \,\rm{Mpc}$, the orders align well with the numerically obtained results. The Laser Interferometer Space Antenna (LISA) is capable of detecting GW signals with amplitudes around $10^{-21}$ and frequencies ranging from $10^{-4}$ to $10^{-1}$ Hz (\cite{Dan2003}). Therefore, GW signals from tidal disruptions involving WDs and IMBHs under our parameter set could be detectable if the source is within a distance of about 1 Mpc from the observer by LISA. Other interferometers like TianQin (\cite{Luo2016}) and DECIGO (\cite{Sato2017}) are also expected to detect GW signals within the frequency range of $10^{-4}$ to $10^{-2}$ Hz. For further details on the prospects of detecting GW signals from tidal disruption events involving WDs and IMBHs, we refer to the discussion in \cite{Toscani2020}.

\section{Discussions and conclusions}
\label{sec4}

To conclude this paper, let us first tie up a few loose ends that we left in section \ref{sec2}. 
First, we note that as mentioned in the beginning of section \ref{sec2}, we have treated the hydrodynamics of the star in a Newtonian manner. 
This approximation can be justified if two conditions are satisfied. (a) The stellar gravity is close to Newtonian and (b) the external gravity
(that due to the BH) varies slowly over scales comparable to the diameter of the star. Both these conditions are satisfied for the IMBH-WD
system that we consider. As for point (a), recall that a useful measure to glean whether a Newtonian treatment of stellar gravity is consistent is given 
by the surface redshift factor $z$, defined via $z+ 1 = (1-2GM_{\star}/c^2R_{\star})^{-1/2}$. Now, for the WD that we have considered, 
$z \simeq 3.5\times 10^{-5}$, which is small enough to indicate that the star is effectively Newtonian (compare this for example with $z \simeq 2\times 10^{-6}$ for the Sun and $z \sim 0.2$ for a typical $1.4M_{\odot}$ neutron star with a radius of 10 Km). One can alternatively consider
the stellar compactness parameter defined as $C = GM_{\star}/(c^2R_{\star})$, see e.g., \cite{ShapiroTeukolsky}. 
In our case $C\sim 3.6\times 10^{-5}$, which is about one order of magnitude less than that of the Sun with $C\sim2.1\times 10^{-6}$.
This indicates that our WD is sufficiently non-compact so that Newtonian hydrodynamics is a good approximation,
compared for example with a typical neutron star of mass $1.4M_{\odot}$ and radius 10 Km for which $C\sim2\times 10^{-1}$.
To see if point (b) is satisfied here,
we note that one often defines a local radius of curvature ${\mathcal R} \sim (r^3c^2/GM)^{1/2}$ to estimate the effect of gravity over a given length 
scale \citep{Taylor, Poisson1, Tejeda2017}. In our case, at $r_p=39 r_g$ where this radius of curvature is minimum, we find ${\mathcal R}/R_{\star} \sim 277$ 
indicating that the effects of BH gravity is indeed a constant to a good approximation over the dimension of the star. Indeed, for closer encounters
with smaller values of $r_p$, this ratio becomes smaller, and the treatment of the stellar fluid using Newtonian hydrodynamics is more challenged. In that
situation, one should possibly write the fluid equations in terms of the Boyer-Lindquist coordinates as done by \cite{Tejeda2017}, which we have not done here. 

The second assumption that we have made is regarding the geodesic nature of the spinning particles that we have considered here (see the discussion
after Equation \ref{Eq.geodesic1}). Now, the deviation from geodesic trajectories of spinning particles can be estimated by converting the 
specific angular momentum of the star to a length scale (upon dividing by $c$) and comparing it with the BH mass on the same scale
(upon multiplying by $G/c^2$), see \cite{Semerak}. The smallness
of the ratio of the specific angular momentum to the BH mass on the same length scale is an indicator of deviation from geodesic behaviour. In our case, we 
numerically computed this ratio at the initial stellar position of $5r_t$ and at $r_p$, and found this to be $1.5\times 10^{-4}$ and $1.3\times 10^{-3}$,
respectively. This indicates that geodesics are a good approximation in the context of the spinning WD that we have considered, see the
discussion in \cite{Semerak}. 

Now we summarize our main results. In this paper, we have considered partial tidal disruptions of spinning WDs in the background of spinning IMBHs in eccentric orbits. We have shown that the ``coupling'' between stellar and BH spins might considerably alter the physics of the process. We observe that the mass disruption from the initial WD depends significantly on both BH and WD rotations, with an approximately $35\%$ relative change between the extreme combinations compared to the case when one considers only stellar or black hole spin. As a result, the mass of the bound debris depends on the coupling of BH and WD spin, and it is approximately the initial WD mass minus the self-bound mass for the parameters we have used, although in some cases, a very small amount of unbound debris is present. Additionally, we find the mass difference between two tidal tails is also a function of the coupled spin. We find that tidal interaction induces rotation on the surviving core. However, when considering both the BH and the WD spin, we find that the profiles are predominantly influenced by the initial spin of the WD. We also note that after the first pericenter passage, the spin angular momenta of both prograde and retrograde spinning WDs are almost identical to the non-spinning ones.

We consider only a single interaction between the WD and the IMBH in an eccentric orbit, resulting in two tidal tails, both of which remain bound to the black hole. This is evident from the differential mass distribution of debris with specific energy figures. The clear separation between the two arms of the mass distribution is a distinct feature compared to parabolic encounters, where only one tail remains bound to the black hole. When the rotations of the BH and the WD are combined, specific characteristics appear in the debris mass distribution plots. We observe that the difference in the initial rise (around $\epsilon/\Delta\epsilon = -2.5$) and the final fall (around $\epsilon/\Delta\epsilon = 0.0$) is mostly influenced by the initial rotation of the WD, as these parts are similar for different BH spins. The mass in the two tails is affected by both the BH and WD spins. For example, with $a^\star = 0.98$ and $\lambda = -0.15$, the two peaks in the mass distribution are amplified (around $\epsilon/\Delta\epsilon = -1.8$ and -0.5). In the mass fallback rate plots, we also observe a dip for all combinations of BH and WD rotation parameters. This behavior is similar to what \cite{Chen2023} found in their study of elliptical partial disruptions with different orbital parameters (see their Figure 5). On the other hand, the most bound debris fallback time varies with different values of $a^\star$ and $\lambda$, primarily due to the initial rotation of the WD. Coupled combinations of spin with more mass result in a wider spread in the fallback plot. Additionally, there is a noticeable change in the late-time fallback rates for different combinations, which is mainly influenced by the WD rotation. With only BH rotation, the late-time fallback rates are almost the same and indistinguishable.

Finally, we discuss the potential observational prospects and future directions related to our work. We have extensively studied the mass fallback rates, which are expected to mimic light curves, as well as the gravitational wave signals, to find out the possible observational signatures due to coupled rotation. We find that coupled rotation can influence the fallback rate, such as by ``broadening'' of 
the fallback curves due to increased debris mass and amplifying the second peak (see the right panels of Figures \ref{fig.RBH1}, \ref{fig.RWD}, \ref{fig.RBHRWD1}), although the effects are weak. However, distinguishing the presence of black hole spin from the most bound debris return time and late-time fallback rate change is challenging, as similar behaviour is produced for both coupled 
BH-WD rotation and spinning WDs. On the other hand, GW signals are less affected by WD spin, making it difficult to infer spin-related details from these observations. While our current findings indicate that observable signatures due to coupled rotation are not large in the parameter regime that we consider, 
more detailed analyses considering some extreme scenarios could provide deeper insights. 
For example, while closer pericenter distances to the event horizon may not significantly amplify the observable effects due to relativistic influences, increasing the stellar rotation could markedly alter the results presented in this paper. This is particularly relevant in scenarios where the star has been spun up to near break-up velocities, which could occur in binary systems through repeated close encounters, as described by \cite{Coughlin2017}.
Further studies in this direction, possibly incorporating general relativistic fluid dynamics within the numerical framework, are essential to deciphering the effects of rotation on observables in these extreme cases.

\section{Acknowledgement}

We thank our anonymous referee for constructive criticisms and suggestions, which significantly improved a draft version of this paper.
We acknowledge the support and resources provided by PARAM Sanganak under the National Supercomputing Mission, Government of India, at the Indian Institute of Technology Kanpur. We thank Kimitake Hayasaki and Pritam Banerjee for very useful comments and discussions. The work of DG is supported by grant number 09/092(1025)/2019-EMR-I from the Council of Scientific and Industrial Research (CSIR). The work of TS is supported in part by the USV Chair Professor position at IIT Kanpur, India. 

\section{Data Availability Statement}

The data underlying this article will be shared upon reasonable request to the corresponding author.







\newpage

\appendix
\section{Code Validation Tests}
\label{app:codetest}

To validate the accuracy and robustness of our SPH code, we present a carefully selected suite of standard numerical tests. These benchmarks assess essential components of the numerical implementation, including adiabatic hydrodynamics (Sod shock tube, Sedov blast), fluid instabilities (Kelvin–Helmholtz instability), and self-gravitating collapse (Evrard collapse). Additionally, the external force module is verified through orbital motion tests in Kerr spacetime, the results of which are detailed in section \ref{sec2a}. To confirm the reliability of our tidal disruption methodology, we also reproduce results from \citet{Cufari2022}.

While our code has undergone comprehensive testing—including pairing instability, random settling, Einfeldt rarefaction, isothermal colliding flows, radial oscillations, and more—we limit this appendix to the most relevant subset of tests to avoid redundancy and maintain clarity. Each selected test specifically addresses a distinct physical aspect critical to our simulations. A complete presentation of all validation exercises will be provided in a future dedicated code methodology paper.

Unless otherwise noted, all tests presented here utilize global time stepping, the $M_6$ quintic spline kernel, and a smoothing length factor $h_{\text{fac}} = 1.0$, where $h_{\text{fac}}$ relates the smoothing length to the mean particle spacing. For gravitational interactions, we employ a tree opening angle of $\theta = 0.5$, ensuring relative acceleration errors remain below 0.1\%. The artificial viscosity parameters are set as $\alpha_{\text{AV}} = 1.0$ and $\beta_{\text{AV}} = 2.0$, while the artificial conductivity parameter is typically $\alpha_u = 1.0$, increased to $\alpha_u = 2.0$ for improved shock resolution in the Sedov blast test.

\subsection*{A1. Sod Shock Tube}

We perform a 3D Sod shock tube test to verify the code’s ability to capture shocks and contact discontinuities. The initial setup consists of two regions separated along the $x$-axis, with densities $\rho_{\text{left}} = 1.0$, $\rho_{\text{right}} = 0.125$ and pressures $P_{\text{left}} = 1.0$, $P_{\text{right}} = 0.1$, with the discontinuity placed at $x = 0$. We use $N \approx 1.5 \times 10^5$ equal-mass particles, distributed on either side of the discontinuity to reproduce the initial density profile. The simulation employs an adiabatic EOS with $\gamma = 5/3$, and periodic boundaries in the $y$ and $z$ directions. In the $x$-direction, particles near the domain boundaries are held fixed to prevent edge effects.

\begin{figure*}
	\centering
	\includegraphics[scale=0.8]{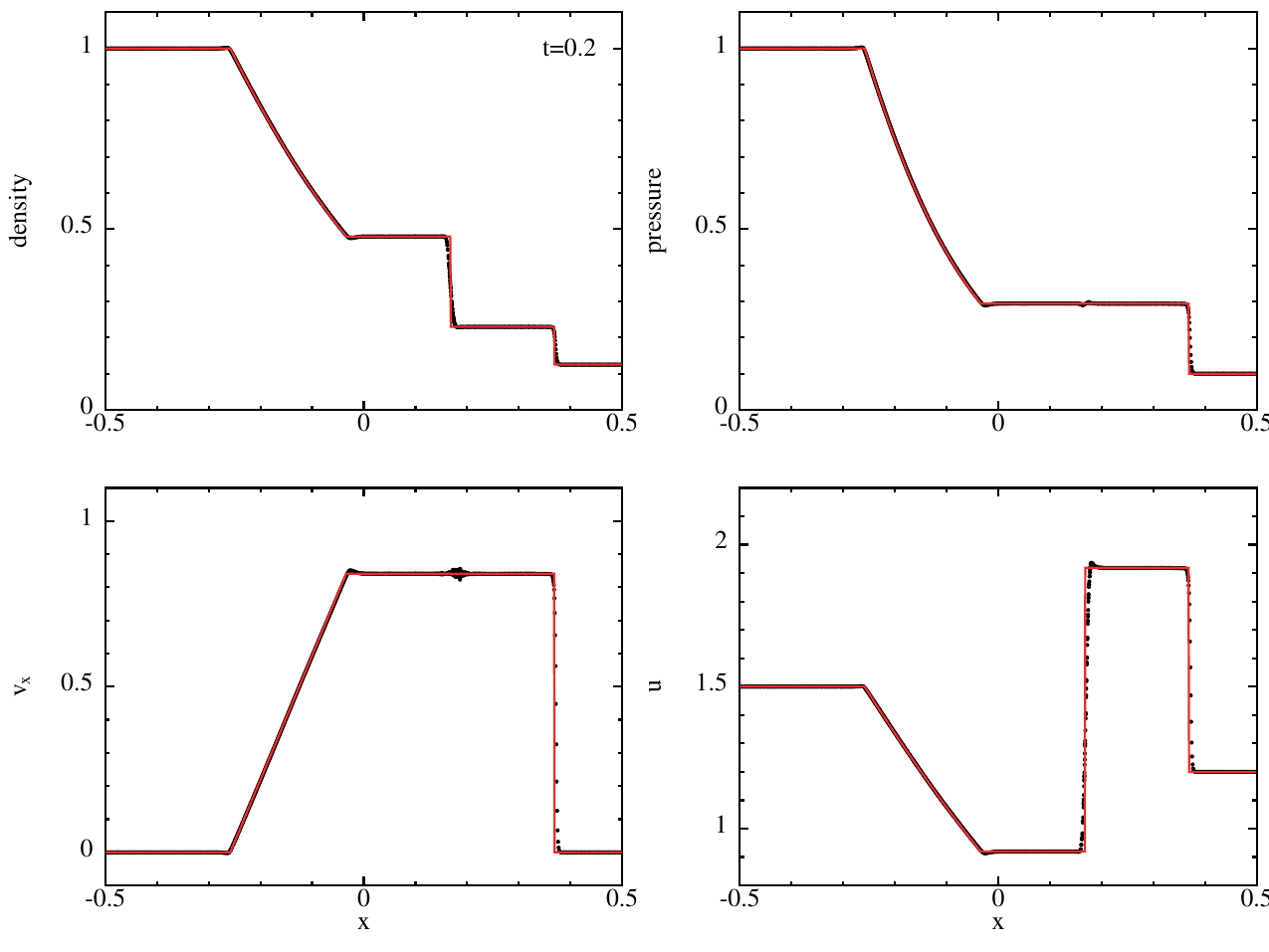}
	\caption{3D shock tube test at $t = 0.2$. Black points represent our SPH results, and the red curves denote the exact solution. The plots were generated using \textsc{SPLASH}.}
	\label{3Dshocktube}
\end{figure*}

Figure~\ref{3Dshocktube} shows the resulting density, pressure, velocity, and internal energy profiles at $t = 0.2$, demonstrating excellent agreement with the exact solution.

\subsection*{A2. Sedov Blast}

We perform the Sedov blast test to assess our code’s accuracy in capturing strong spherical shocks. We initialize particles using a ``glass'' distribution, similar to the method adopted in the SEREN code \citep{Hubber}. A glass distribution arranges particles approximately equidistantly, reducing directional biases and lattice artifacts. We assume an adiabatic EOS with $\gamma = 5/3$. Approximately $10^5$ particles are used, with the central particle and its $\sim50$ nearest neighbors receiving a total thermal energy of unity distributed according to the smoothing kernel. The remaining particles have a significantly lower initial thermal energy, $10^{-6}$ times smaller than that of the most energetic (central) particle. This energy configuration produces a spherical shock wave propagating outward, compressing surrounding material into a dense shell.

\begin{figure*}
	\centering		
	\includegraphics[scale=0.3]{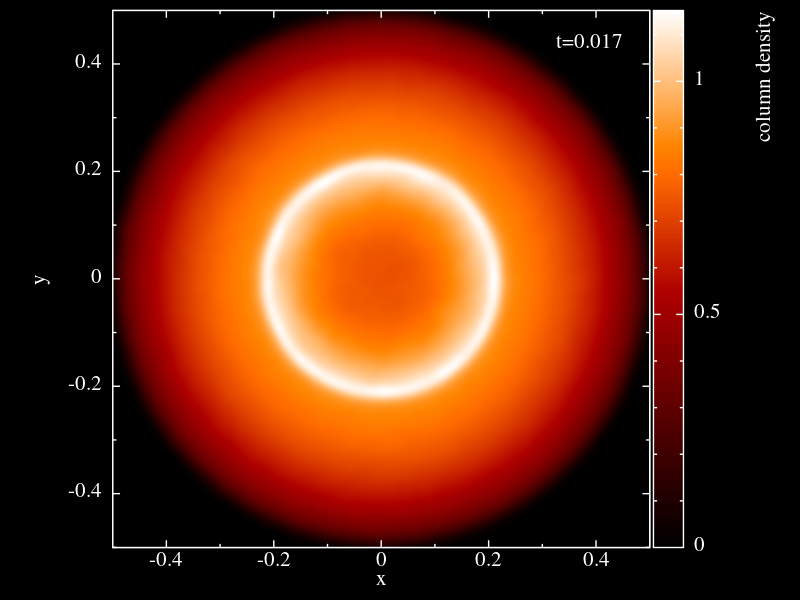}	
	\includegraphics[scale=0.3]{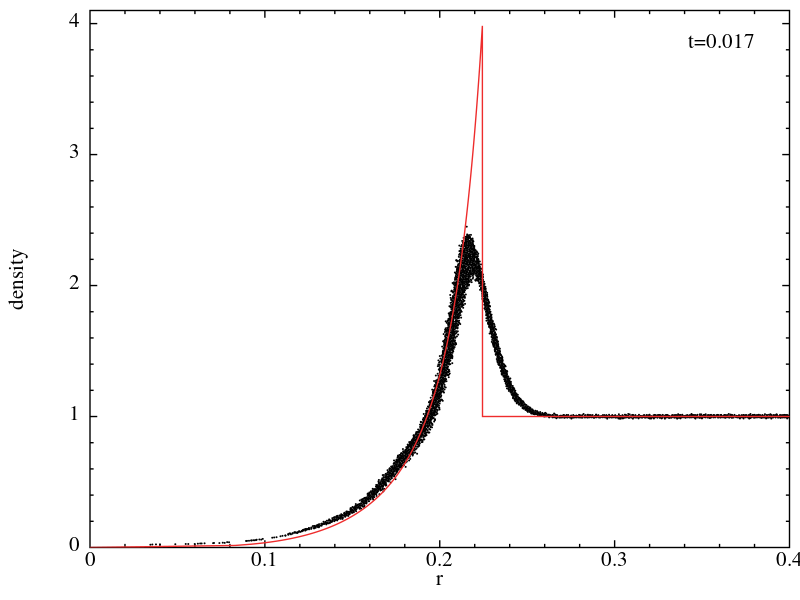}	
	\caption{3D Sedov blast. \textbf{Left panel:} rendered plot of the density and \textbf{right panel:} density profile shows the blast wave propagating outwards at $t=0.017$. Red line is the analytical solution. This figure is plotted in SPLASH.}
	\label{3dsedov}
\end{figure*}

Figure~\ref{3dsedov} shows agreement with the analytical solution.

\subsection*{A3. Kelvin–Helmholtz Instability}

To assess the code’s ability to model shear-driven mixing and instabilities at contact discontinuities, we perform a 3D Kelvin–Helmholtz (KH) instability test using $N \approx 4.4\times10^5$ particles. The setup is effectively two-dimensional, with densities $\rho = 2$ in the region $\vert y \vert \leq 0.25$ and $\rho = 1$ elsewhere, pressure equilibrium at $P = 2.5$, and an adiabatic EOS with $\gamma = 5/3$. The density profile is generated via stretch mapping along the $y$-direction. Initial velocities are $v_x = 0.5$ for $\vert y \vert \leq 0.25$ and $v_x = -0.5$ outside this region. A small sinusoidal velocity perturbation following the parameters of the K1 run in \citet{Merlin} is applied along the $y$-direction to trigger the instability.

\begin{figure*}
	\centering		
	\includegraphics[scale=0.4]{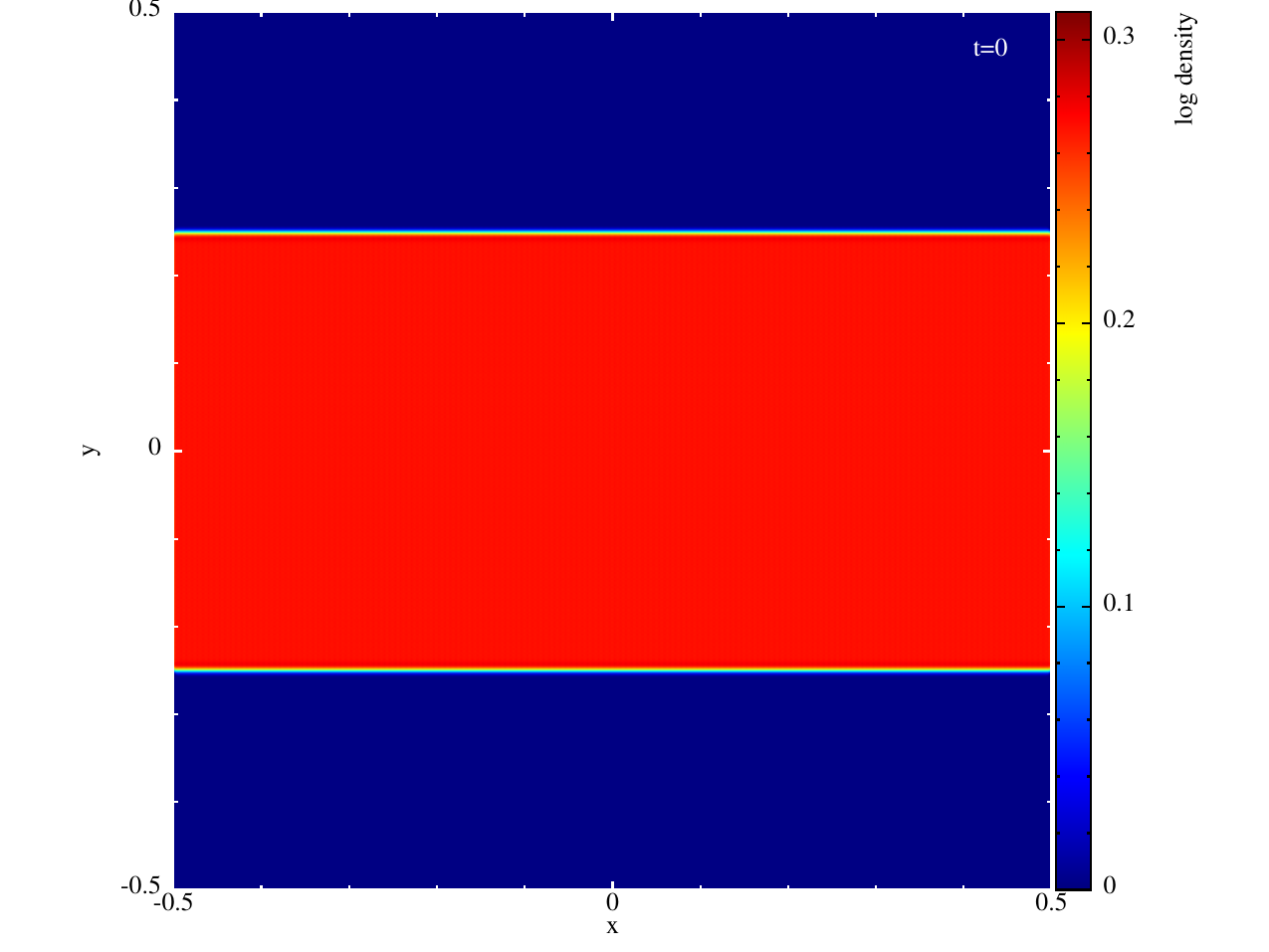}
	\includegraphics[scale=0.4]{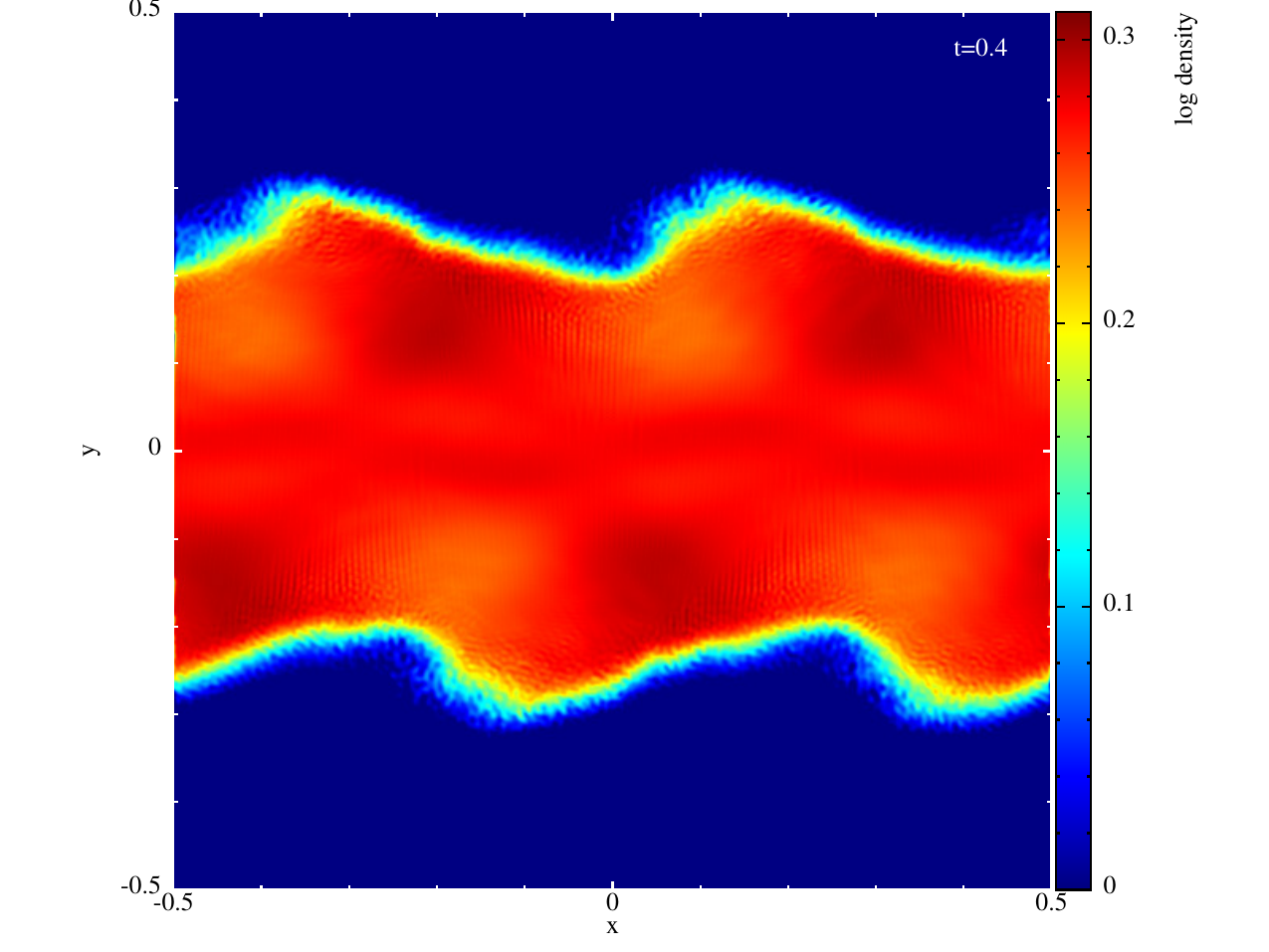}
	\includegraphics[scale=0.4]{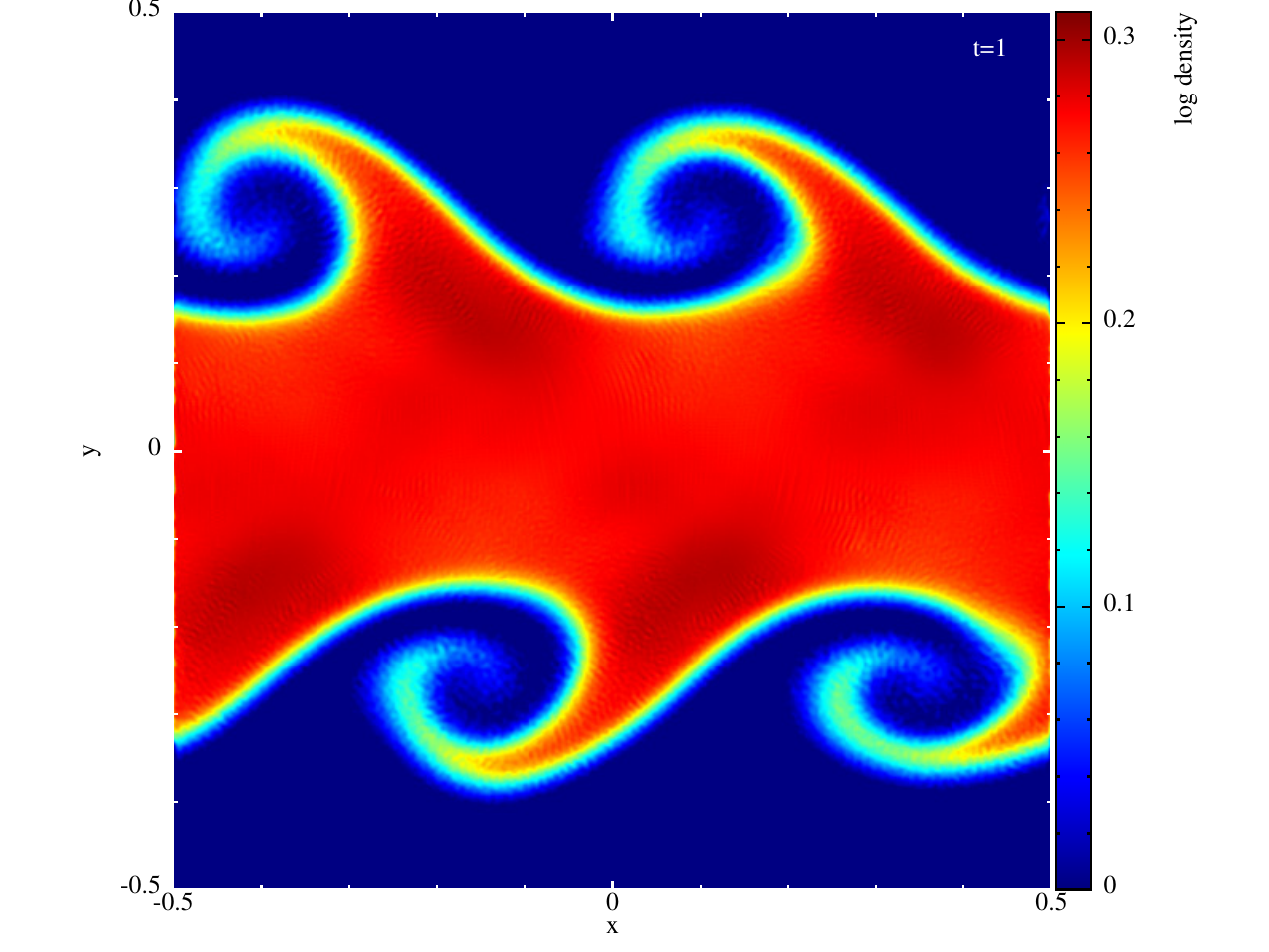}
	\includegraphics[scale=0.4]{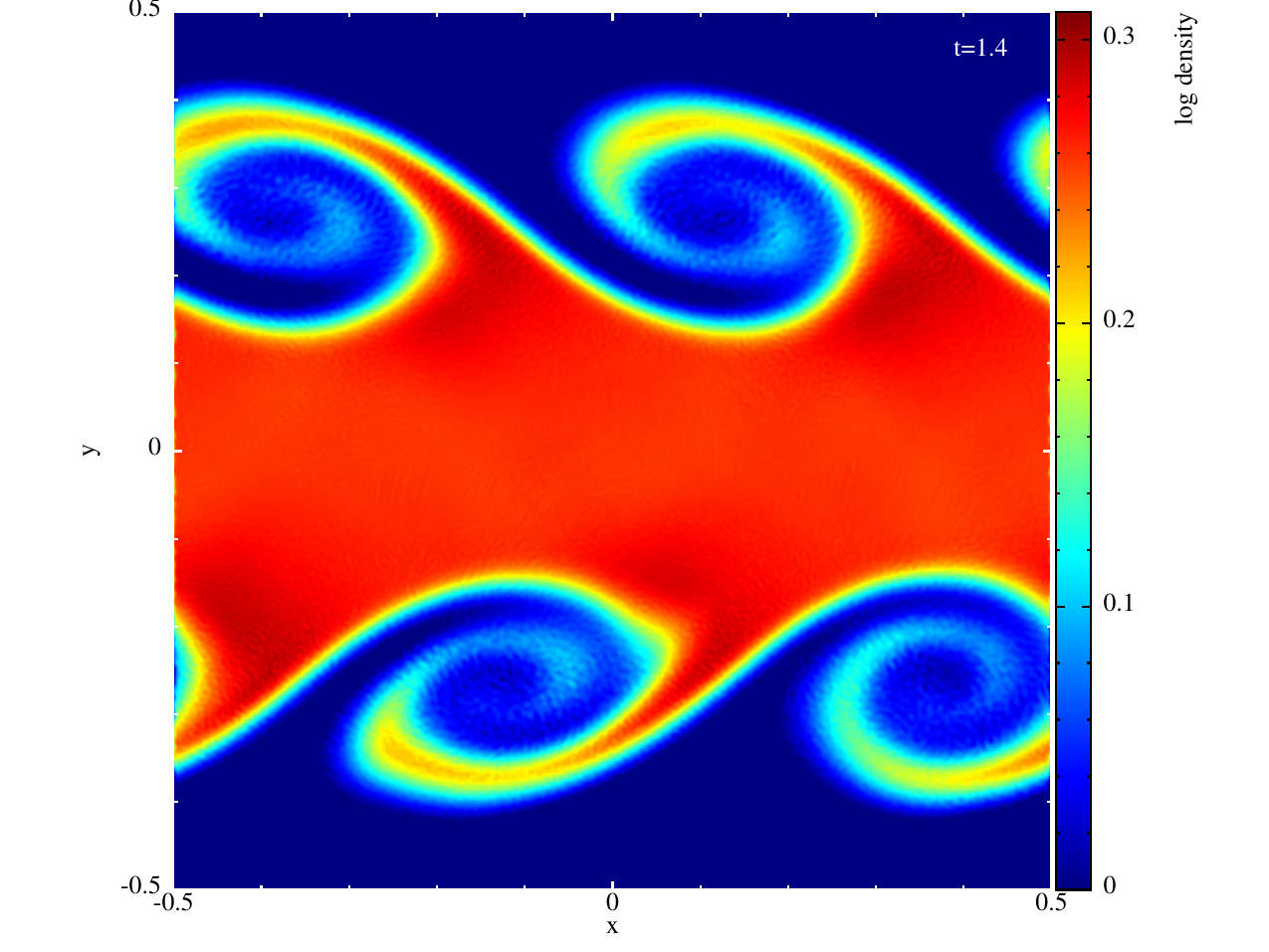}		
	\caption{Snapshots from the 3D Kelvin–Helmholtz instability test showing density in the $z=0$ plane at $t = 0.0$, $0.4$, $1.0$, and $1.4$. The initial equilibrium evolves under the applied velocity perturbation, producing the characteristic KH rolls. Clearly, the code reproduces the expected formation of whirlpools and mixing between the fluid layers. Figures are rendered using SPLASH.}
	\label{3DKH}
\end{figure*}

Figure~\ref{3DKH} shows density snapshots in the $z=0$ plane at $t = 0.0$, $0.4$, $1.0$, and $1.4$, capturing the initial equilibrium, onset, and growth of the instability. Our code reproduces the expected development of whirlpools and progressive mixing between the two fluid layers as the KH rolls form and evolve.

\subsection*{A4. Evrard Collapse}

We perform the Evrard collapse to validate self-gravity, shock handling, and energy conservation,  at two resolutions ($N \approx 10^{5}$ and $10^{6}$ particles). The initially static sphere of mass $M = 1$ and radius $R = 1$ follows the density profile $\rho(r)=M/(2\pi R^{2}r)$ (generated via stretch mapping) and has specific internal energy $u = 0.05\,GM/R$; an adiabatic EOS with $\gamma = 5/3$ is used.

\begin{figure*}
	\centering		
	\includegraphics[scale=0.5]{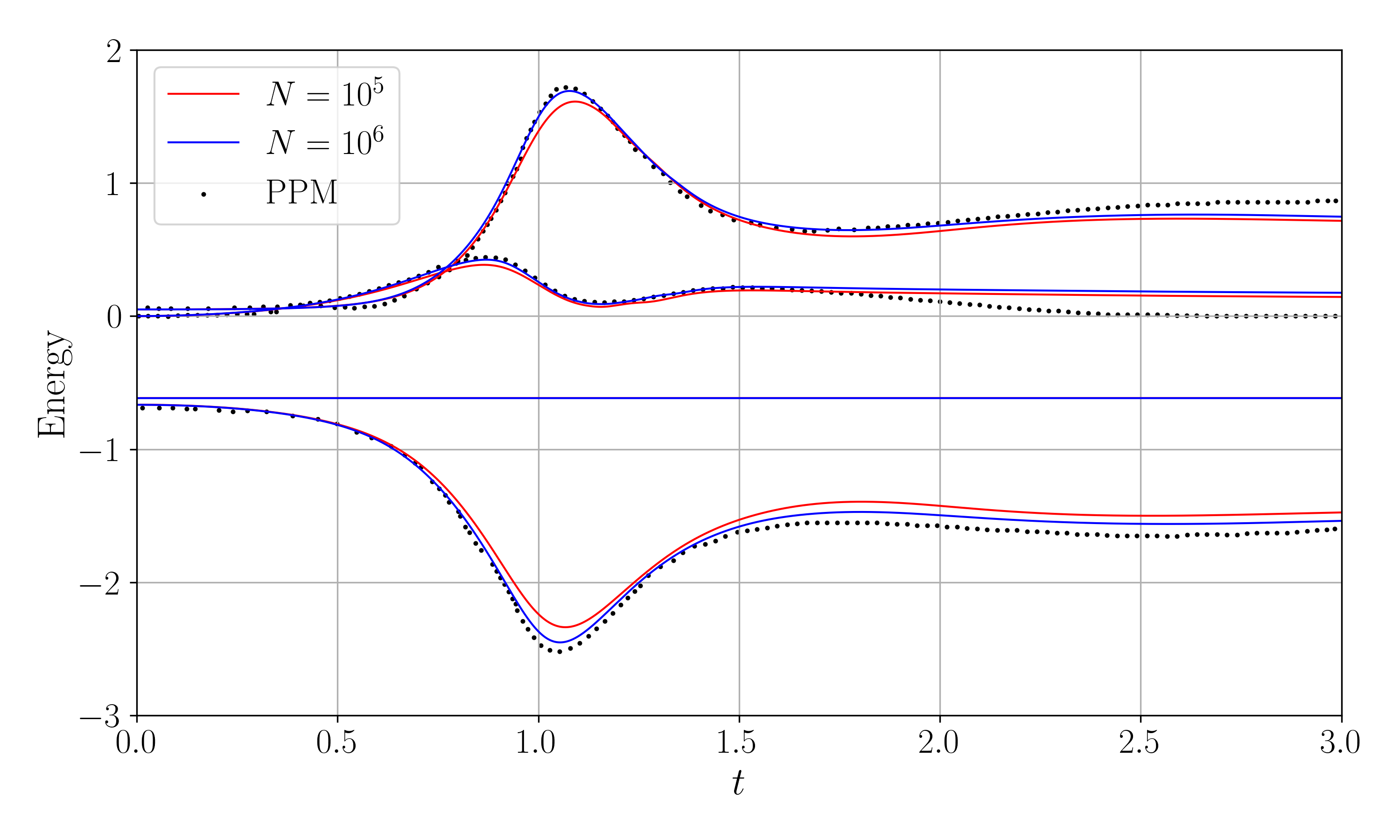}	
	\caption{Evolution of thermal, kinetic, total, and potential energies during the Evrard collapse. Black points show the 1D PPM reference solution taken from Figure~6 of \citet{Stein}, while the coloured curves correspond to our SPH runs at different resolutions.}
	\label{3devrard2}
\end{figure*}

Figure~\ref{3devrard2} plots kinetic, thermal, potential, and total energies to $t = 3$, together with the 1D piecewise parabolic method (PPM) reference solution of \citet{Stein}. Up to $t \lesssim 1.5$, both SPH resolutions track the PPM energies well within the limit. After the rebound ($ t\gtrsim 2$), the SPH energies converge toward values slightly offset from the PPM curve -- behaviour also reported in the PHANTOM comparison study. However, total energy remains conserved within $0.02\%$ at $t = 3$ for both resolutions. 

\begin{figure*}
	\centering		
	\includegraphics[scale=0.34]{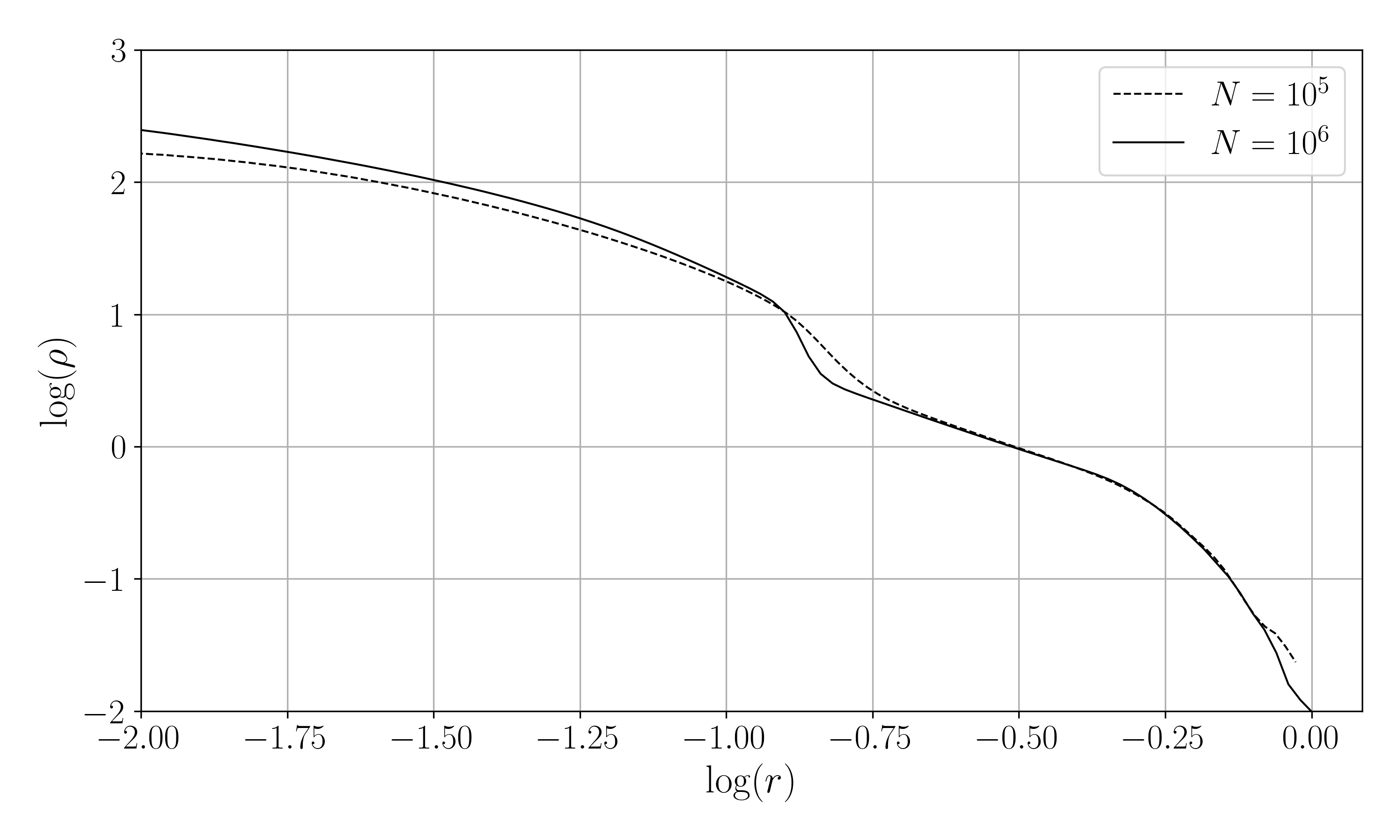}
	\includegraphics[scale=0.34]{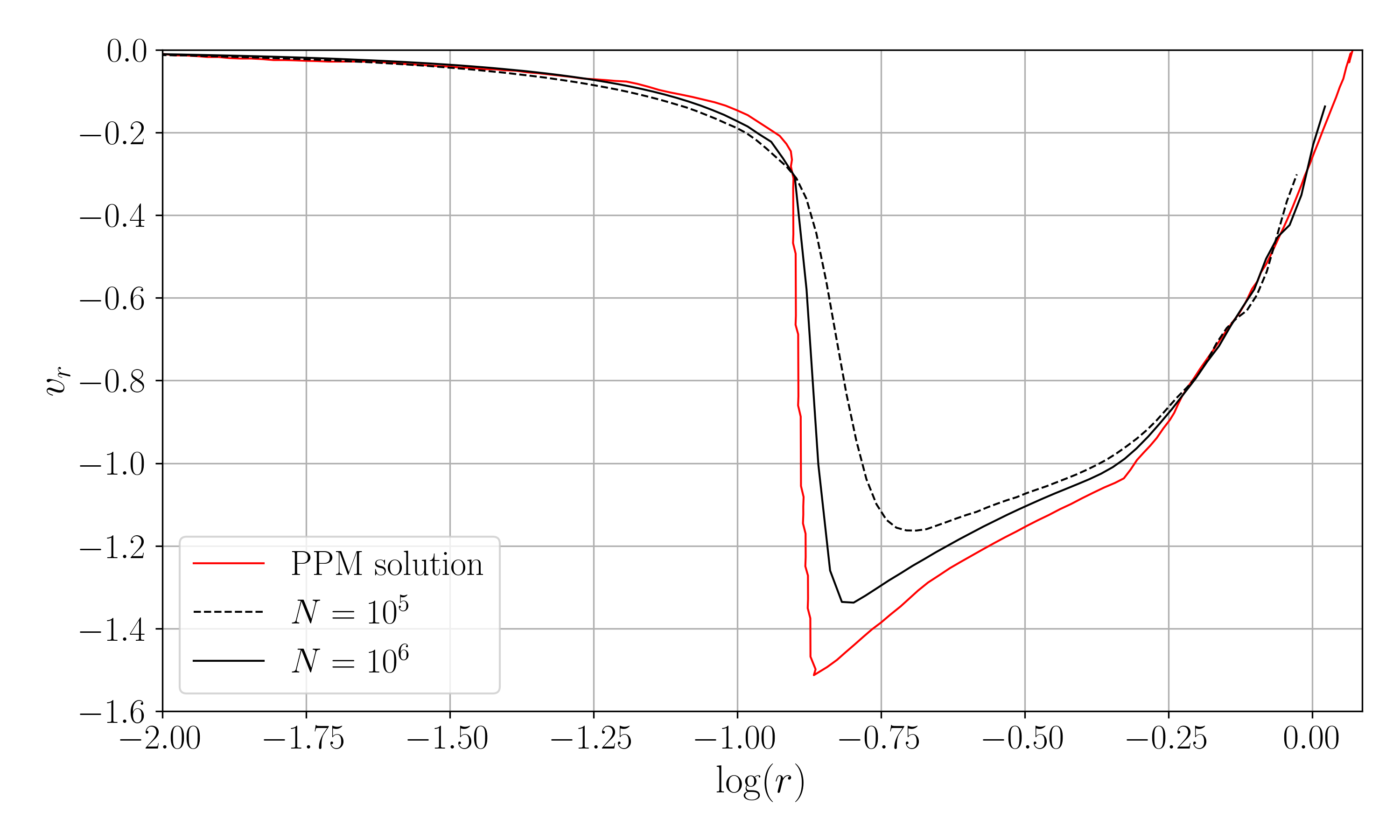}	
	\caption{Radial profiles from the 3D Evrard collapse at $t = 0.77$. \textbf{Left panel:} logarithmic density, $\log\rho(r)$, \textbf{Right panel:} radial velocity, $v_{r}(r)$, both plotted against $\log r$. Black solid curve: high-resolution run ($N \approx 10^{6}$); black dashed curve: lower-resolution run ($N \approx 10^{5}$). The red curve shows the 1D PPM reference solution taken from Figure 7 of \citet{Stein}. The outward-propagating shock at $r \simeq 0.1$ is increasingly sharp and well aligned with the reference solution as the resolution is raised, demonstrating convergence.}
	\label{3devrard1}
\end{figure*}

Figure~\ref{3devrard1} shows radial profiles at $t = 0.77$, when the outward-moving shock lies at $r \simeq 0.1$. The left panel (density) and right panel (radial velocity) demonstrate excellent agreement between the different resolution runs and close correspondence with the PPM solution, confirming that the shock location and post-shock structure are accurately captured.

Together, the energy and radial profiles verify that our artificial-viscosity parameters reproduce the correct collapse, bounce, and shock propagation without introducing excessive dissipation.

\subsection*{A5. Comparison with PHANTOM Benchmark}

To verify our tidal-disruption implementation, we replicate the setup of Figure 4 (top-left panel) in \citet{Cufari2022}. Both studies follow the full disruption of a $\gamma = 5/3$ polytropic star on an eccentric orbit ($e \simeq 0.9$–0.93) about a black hole. Using identical initial conditions, our simulation reproduces the key diagnostics reported for PHANTOM -- most importantly the fallback rate $\dot{M}(t)$. 

\begin{figure*}
	\centering		
	\includegraphics[scale=0.3]{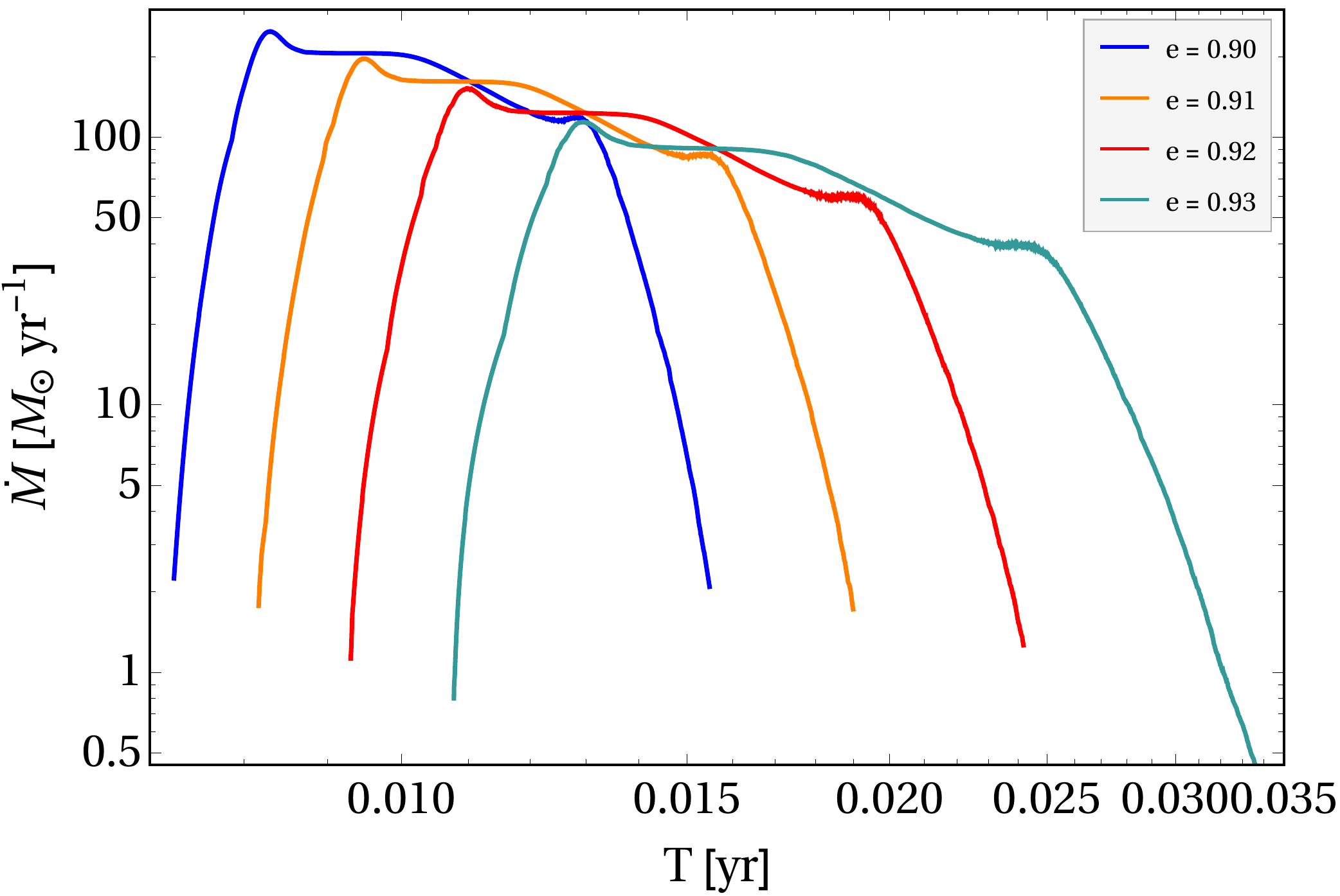}	
	\caption{Fallback rate ${\dot M}(t)$ from our code for the same $\gamma=5/3$ polytropic disruption simulated in Figure 4 (top-left) of \citet{Cufari2022}. The curve reproduces the amplitude and temporal evolution reported in the PHANTOM calculation, demonstrating that our TDE methodology captures the correct accretion behaviour.}
	\label{nixon}
\end{figure*}

Figure \ref{nixon} plots our $\dot{M}(t)$ curve, showing that its amplitude and temporal evolution agree closely with the PHANTOM result. This correspondence confirms that our treatment of hydrodynamics, self-gravity, and debris morphology yields outcomes consistent with the established benchmark.

\section{Convergence Test}
\label{app:convergence}

\begin{figure*}
	\centering		
	\includegraphics[scale=0.25]{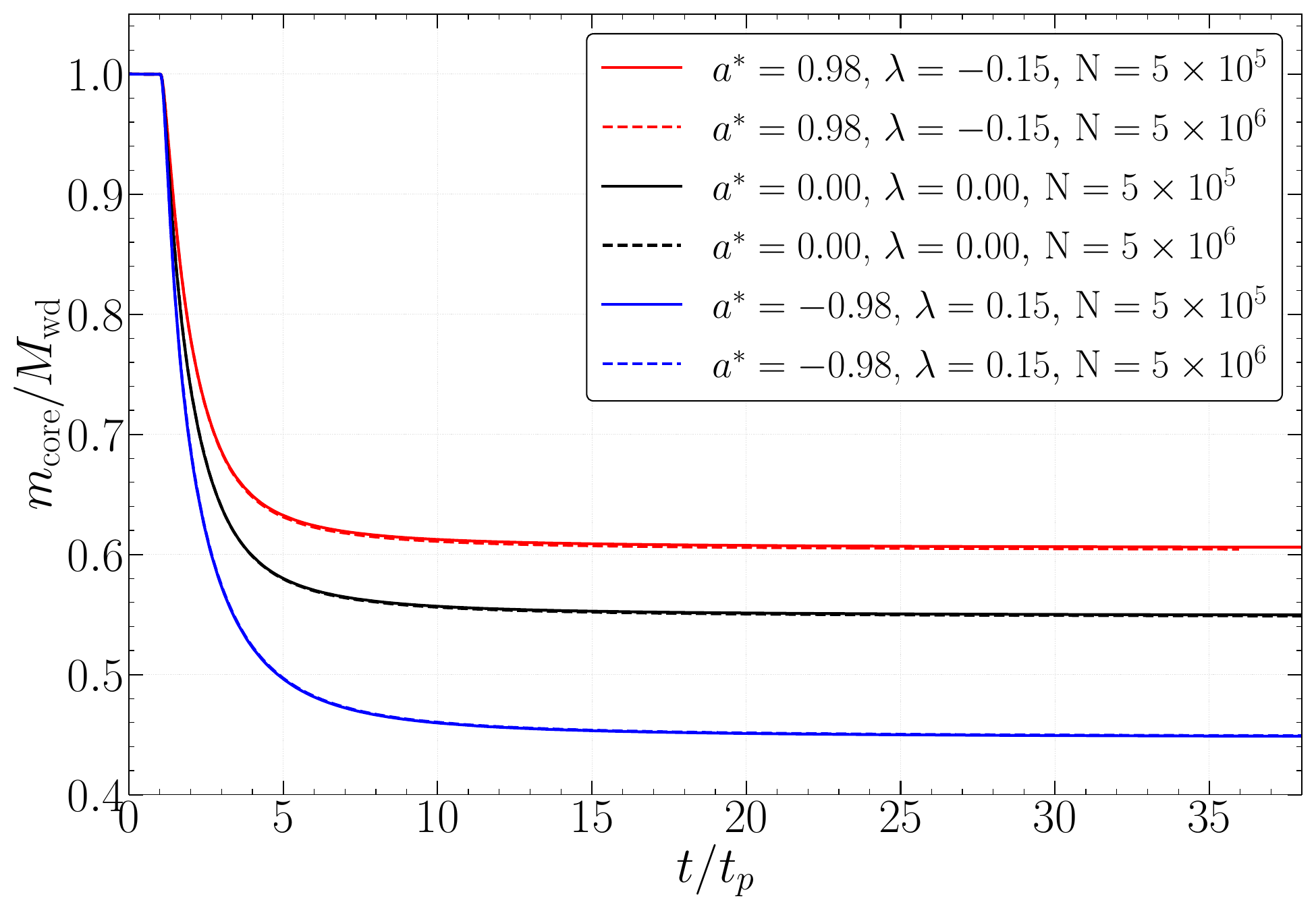}	
	\includegraphics[scale=0.25]{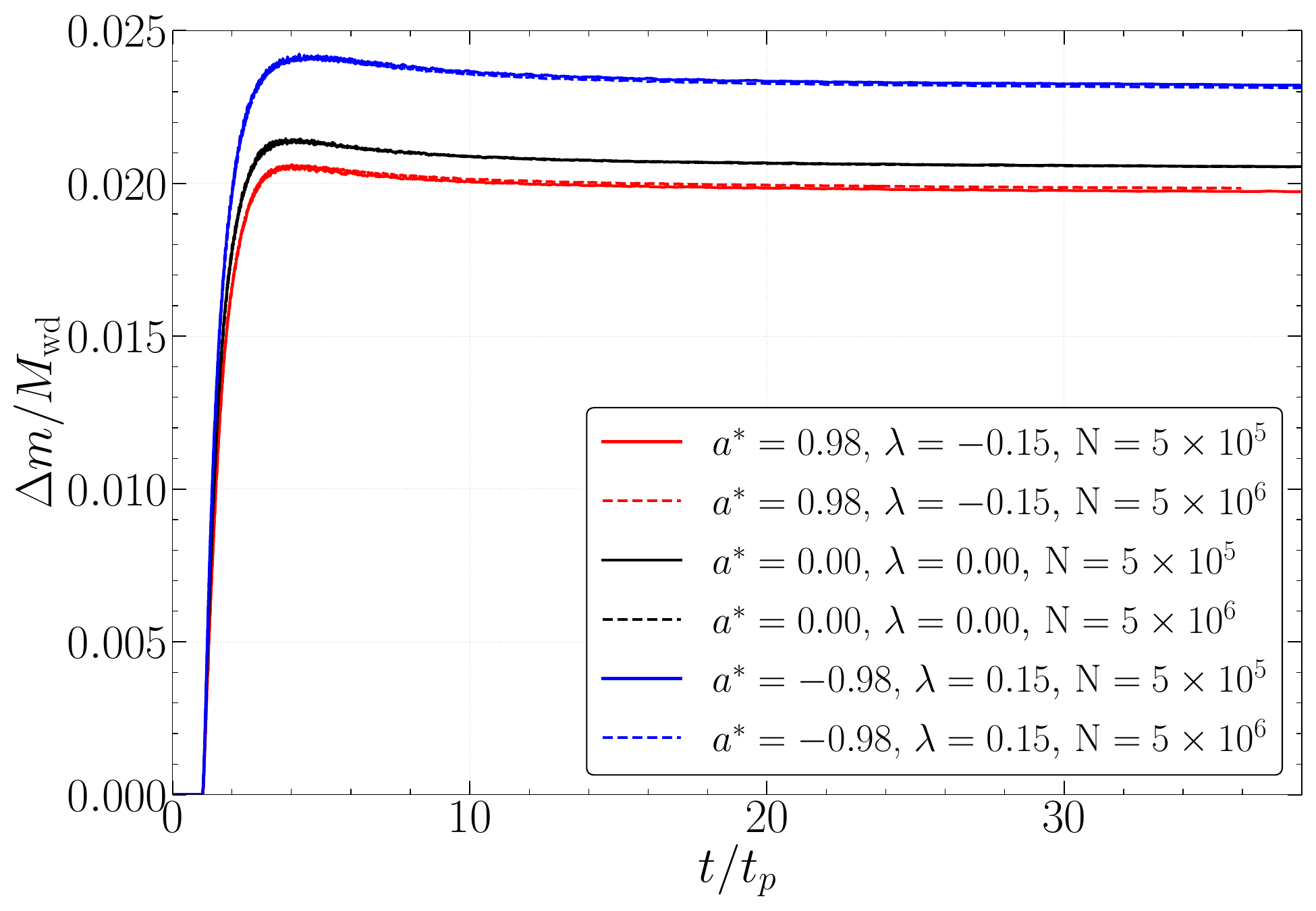}	
	\includegraphics[scale=0.25]{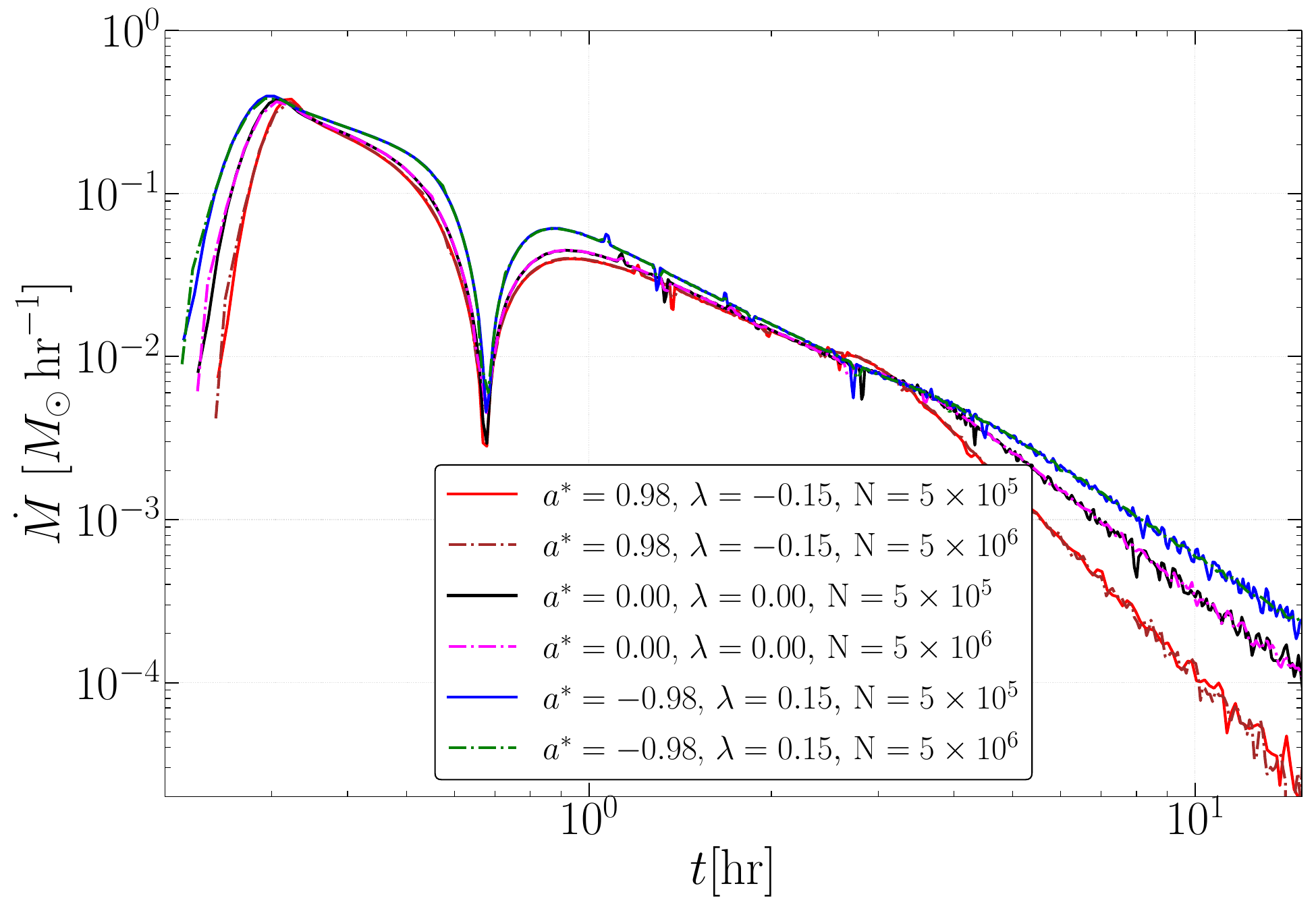}	
	\includegraphics[scale=0.25]{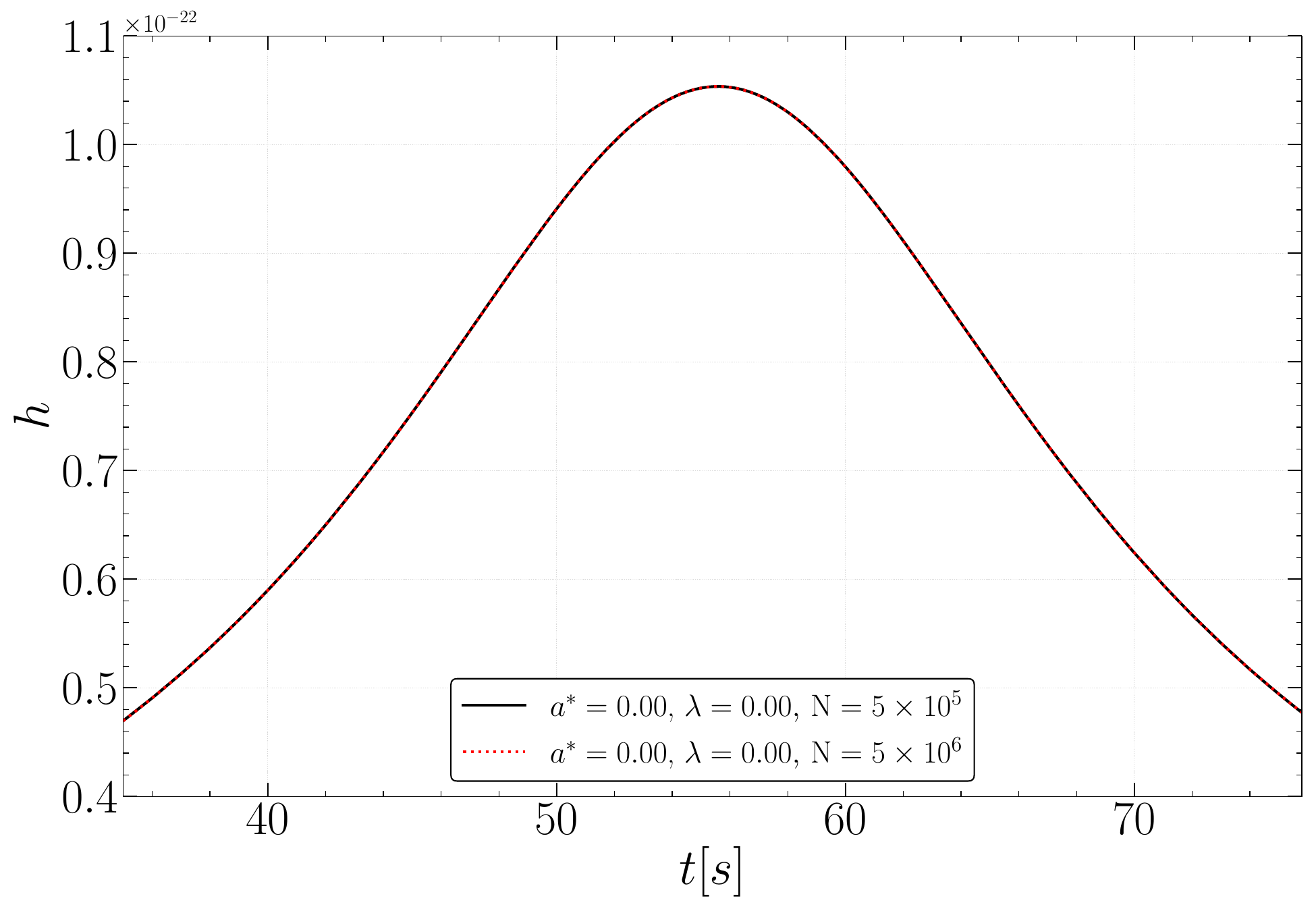}
	\caption{We compare the core mass (\textbf{Top left}), the mass difference between two tails (\textbf{Top right}), fallback rates of the bound debris (\textbf{Bottom left}), and gravitational wave amplitude variations (\textbf{Bottom right}) for the two resolutions mentioned in the legend. Both resolutions provide identical results, validating our resolution choice. }
	\label{fig:appendix_convergence}
\end{figure*}

To assess the sensitivity of our results to numerical resolution, we performed a convergence test by comparing simulations with $5\times 10^5$ and $5\times 10^6$ SPH particles for a few representative spin configurations. Figure~\ref{fig:appendix_convergence} shows that the key quantities remain consistent between the two resolutions, confirming that our fiducial choice is sufficient.

\section{Evolution of Individual Tidal Tail Masses}
\label{app:m1m2}

\begin{figure*}
	\centering		
	\includegraphics[scale=0.25]{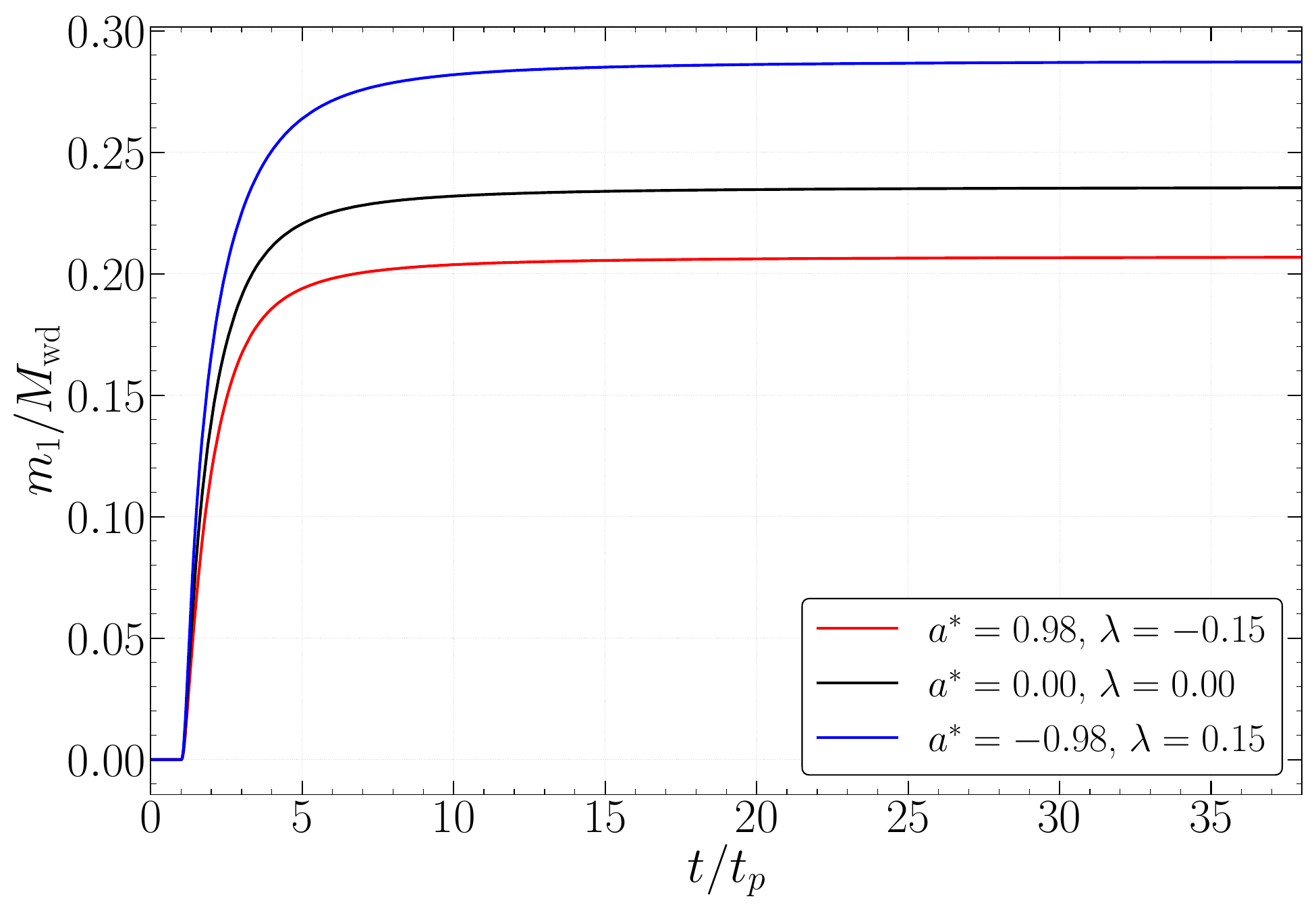}	
	\includegraphics[scale=0.25]{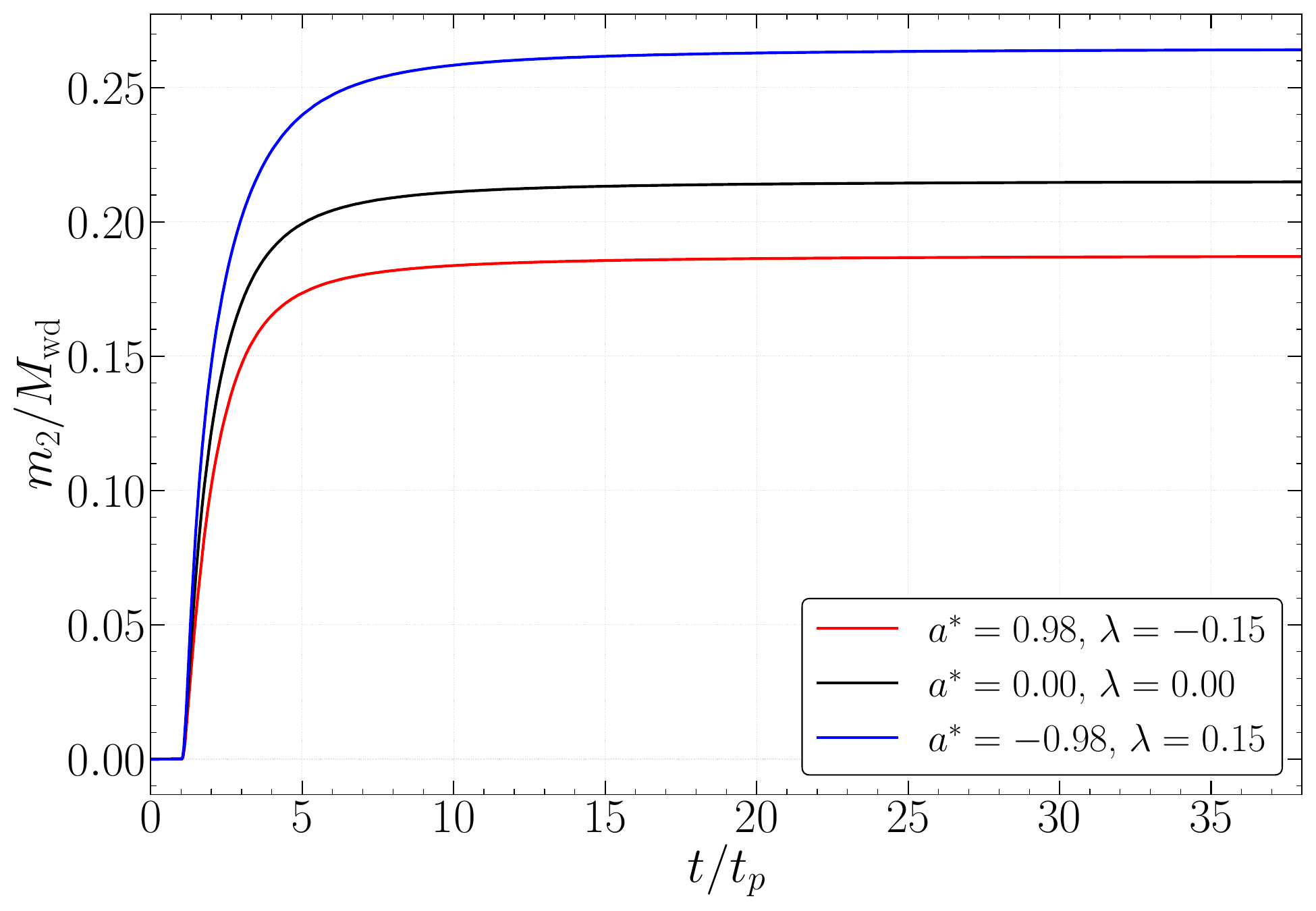}	
	\caption{\textbf{Left panel:} Normalized mass of the tidal tail closest to the BH, $m_1/M_{\rm wd}$, as a function of time in units of $t_p$. \textbf{Right panel:} Same for the tail farther from the BH, $m_2/M_{\rm wd}$.}
	\label{fig:appendix_tailmass}
\end{figure*}

To complement the analysis in the main text, we present the time evolution of the individual tail masses $m_1$ and $m_2$ in Figure~\ref{fig:appendix_tailmass}. As defined, $m_1$ is the mass of the tidal tail closer to the BH, and $m_2$ is the mass of the tail farther from the BH. These quantities are normalized by the initial WD mass and are shown for the spin combinations mentioned in the legend. The divergence between $m_1$ and $m_2$ over time illustrates the origin of the asymmetry leading to the $\Delta m$ trend discussed in the right panel of Figure~\ref{fig.RBHRWD}.

\bsp	
\label{lastpage}
\end{document}